\documentclass{aa}  

\usepackage{graphicx}
\usepackage{txfonts}
\usepackage{float}
\usepackage{amsmath}
\usepackage[usenames,dvipsnames]{color}
\usepackage[colorlinks=false, linktocpage=true]{hyperref}

\usepackage{siunitx}

\newcommand{\kpc}{{\ifmmode{\mathrm{kpc}}\else{$\mathrm{kpc}$\xspace}\fi}}
\newcommand{\Mpc}{{\ifmmode{\mathrm{Mpc}}\else{$\mathrm{Mpc}$\xspace}\fi}}
\newcommand{\Gpc}{{\ifmmode{\mathrm{Gpc}}\else{$\mathrm{Gpc}$\xspace}\fi}}

\newcommand{\hkpc}{{\ifmmode{h^{-1}\mathrm{kpc}}\else{$h^{-1}\mathrm{kpc}$\xspace}\fi}}
\newcommand{\hMpc}{{\ifmmode{h^{-1}\mathrm{Mpc}}\else{$h^{-1}\mathrm{Mpc}$\xspace}\fi}}
\newcommand{\hGpc}{{\ifmmode{h^{-1}\mathrm{Gpc}}\else{$h^{-1}\mathrm{Gpc}$\xspace}\fi}}

\newcommand{\MpcCu}{{\ifmmode{\mathrm{Mpc}^3}\else{$\mathrm{Mpc}^3$\xspace}\fi}}
\newcommand{\MpcV}{{\ifmmode{\mathrm{Mpc}^{-3}}\else{$\mathrm{Mpc}^{-3}$\xspace}\fi}}

\newcommand{\hMsun}{{\ifmmode{h^{-1}\mathrm{M_{\odot}}}\else{$h^{-1}\mathrm{M_{\odot}}$}\fi}}
\newcommand{\Msun}{{\ifmmode{\mathrm{M_{\odot}}}\else{$\mathrm{M_{\odot}}$\xspace}\fi}}

\newcommand{\Msunyr}{{\ifmmode{\mathrm{M_{\odot}yr^{-1}}}\else{$\mathrm{M_{\odot}yr^{-1}}$\xspace}\fi}}
\newcommand{\Gyr}{{\ifmmode{\mathrm{Gyr}}\else{$\mathrm{Gyr}$\xspace}\fi}}
\newcommand{\yr}{{\ifmmode{\mathrm{yr}}\else{$\mathrm{yr}$\xspace}\fi}}
\newcommand{\yrmo}{{\ifmmode{\mathrm{yr}^{-1}}\else{$\mathrm{yr}^{-1}$\xspace}\fi}}

\newcommand{\kms}{{\ifmmode{\mathrm{kms}^{-1}}\else{$\mathrm{kms}^{-1}$\xspace}\fi}}
\newcommand{\Zsolar}{{\ifmmode{\mathrm{Z}_{\odot}}\else{$\mathrm{Z}_{\odot}$\xspace}\fi}}

\newcommand{\SFR}{{\ifmmode{\mathrm{SFR}}\else{$\mathrm{SFR}$\xspace}\fi}}
\newcommand{\SFRs}{{\ifmmode{SFRs}\else{$\mathrm{SFRs}$\xspace}\fi}}
\newcommand{\sSFR}{{\ifmmode{\mathrm{sSFR}}\else{$\mathrm{sSFR}$\xspace}\fi}}
\newcommand{\sSFRs}{{\ifmmode{\mathrm{sSFRs}}\else{$\mathrm{sSFRs}$\xspace}\fi}}

\newcommand{\Mstar}{{\ifmmode{\mathrm{M_{\star}}}\else{$\mathrm{M_{\star}}$\xspace}\fi}}

\newcommand{\Mhalo}{{\ifmmode{\mathrm{M_{200}}}\else{$\mathrm{M_{200}}$\xspace}\fi}}

\begin{document}

\title{Galaxy populations in protoclusters at cosmic noon}

\author{
    Moira Andrews\thanks{moira\_andrews@ucsb.edu}\inst{1, 2, 3}
    \and 
    M. Celeste Artale\thanks{maria.artale@unab.cl}\inst{1, 4}
    \and 
    Ankit Kumar\inst{4}
    \and 
    Kyoung-Soo Lee\inst{1}
    \and 
    Tess Florek\inst{1}
    \and 
    Kaustub Anand\inst{1}
    \and 
    Candela Cerdosino\inst{5}
    \and 
    Robin Ciardullo\inst{6, 7}
    \and 
    Nicole Firestone\inst{8}
    \and
    Eric Gawiser\inst{8} 
    \and 
    Caryl Gronwall\inst{6, 7}
    \and 
    Lucia Guaita\inst{4}
    \and 
    Sungryong Hong\inst{9}
    \and 
    Ho Seong Hwang\inst{10, 11}
    \and 
    Jaehyun Lee\inst{9}
    \and 
    Seong-Kook Lee\inst{10, 11}
    \and
    Nelson Padilla\inst{5}
    \and
    Jaehong Park\inst{12, 15}
    \and 
    Roxana Popescu\inst{13}
    \and 
    Vandana Ramakrishnan\inst{1}
    \and 
    Hyunmi Song\inst{14}
    \and 
    F. Vivanco C\'adiz\inst{4}
    \and
    Mark Vogelsberger\inst{16}
}

\institute{
    Department of Physics and Astronomy, Purdue University, 525 Northwestern Ave., West Lafayette, IN 47907, USA
    \and
    Department of Physics, University of California, Santa Barbara, CA 93106-9530, USA
    \and
    Las Cumbres Observatory, 6740 Cortona Dr, Suite 102, Goleta, CA 93117-5575, USA
    \and
    Universidad Andres Bello, Facultad de Ciencias Exactas, Departamento de Fisica y Astronomia, Instituto de Astrofisica, Fernandez Concha 700, Las Condes, Santiago RM, Chile
    \and
    Instituto de Astronomía Teórica y Experimental (IATE), CONICET-UNC, Laprida 854, X500BGR, Córdoba, Argentina
    \and
    Department of Astronomy \& Astrophysics, The Pennsylvania State University, University Park, PA 16802, USA
    \and
    Institute for Gravitation and the Cosmos, The Pennsylvania State University, University Park, PA 16802, USA
    \and
    Physics and Astronomy Department, Rutgers, The State University, Piscataway, NJ 08854, USA
    \and
    Korea Astronomy and Space Science Institute, 776 Daedeokdae-ro, Yuseong-gu, Daejeon 34055, Korea
    \and
    Department of Physics and Astronomy, Seoul National University, 1 Gwanak-ro, Gwanak-gu, Seoul 08826, Korea
    \and
    SNU Astronomy Research Center, Seoul National University, 1 Gwanak-ro, Gwanak-gu, Seoul 08826, Korea
    \and
    Korea Institute for Advanced Study, 85 Hoegi-ro, Dongdaemun-gu, Seoul 02455, Korea
    \and
    Department of Astronomy, University of Massachusetts, Amherst, MA 01003, USA
    \and
    Department of Astronomy and Space Science, Chungnam National University, 99 Daehak-ro, Yuseong-gu, Daejeon, 34134, Korea
    \and
    Space Science Exploration Directorate, Korea AeroSpace Administration (KASA), 537 Haeansaneop-ro, Sanam-myeon, Sacheon-si, Gyeongsangnam-do, Korea
    \and
    Kavli Institute for Astrophysics and Space Research, Massachusetts Institute of Technology,
    Cambridge, MA 02139, USA
}

   \date{Received 2024; accepted 2025}

  \abstract
   {}
   {We investigate the physical properties and redshift evolution of simulated galaxies residing in unvirialized cosmic structures (i.e., protoclusters) at cosmic noon, to understand the influence of the environment on galaxy formation. This work is intended to build clear expectations for the ongoing ODIN (One-hundred-deg$^2$ DECam Imaging in Narrowbands) survey, which is mapping large-scale structures at $z =2.4, 3.1,$ and 4.5 using Ly$\alpha$-emitting galaxies (LAEs) as tracers.}
   {From the  IllustrisTNG simulations, we define subregions centered on the most massive clusters ranked by total stellar mass at $z=0$ and study the properties of galaxies within, including those of LAEs. To model the LAE population, we take a semi-analytical approach that assigns Ly$\alpha$ luminosity and equivalent width based on the UV luminosities to galaxies in a probabilistic manner.
   We investigate stellar mass, star formation rate (SFR), major merger events, and specific star formation rate of the population of star-forming galaxies and LAEs in the field- and protocluster environment and trace their evolution across cosmic time between $z=0-4$.  }
   {
   We find that the overall shape of the UV luminosity function in simulated protocluster environments is characterized by a substantially shallower faint-end slope and a large excess on the bright end, signaling different formation histories for galaxies therein. The difference is milder for the Ly$\alpha$ luminosity function. 
   While protocluster galaxies follow the same SFR-$\Mstar$ scaling relation as average field galaxies, a larger fraction appears to have experienced major mergers in the last 200~Myr and as a result shows enhanced star formation at a $\approx$60\% level, leading to a flatter distribution in both SFR and $\Mstar$ relative to galaxies in the average field. 
   We find that protocluster galaxies, including LAEs, begin to quench much earlier ($z\sim0.8-1.6$) than field galaxies ($z\sim 0.5-0.9$); our result is in qualitative agreement with recent observational results and highlights the importance of large-scale environment on the overall formation history of galaxies.
   }
   {}
   

   \maketitle

\section{Introduction}
 
 Galaxy clusters are the most massive virialized 
 structures in the present-day Universe. With halo masses exceeding $10^{14}$\,\Msun, these systems host hundreds to thousands of member galaxies, mostly with passively evolving stellar populations \citep{Kravtsov2012,Overzier_2016,Alberts_2022}. Estimated ages and $\alpha$/Fe elemental ratios of cluster galaxies suggest that much of the star formation activity therein occurred at high redshift, $z\gtrsim 2$ \citep{Thomas05}. Thus, exploring the early phases of galaxy clusters is pivotal for understanding how the environment shapes their galaxy populations.

What drives the accelerated formation and fast quenching in high-density environments is poorly understood. This is in part due to the technical challenges of identifying 
these unvirialized protoclusters, which, at $z\gtrsim 2$, span tens of comoving megaparsecs \citep{Chiang_2013,Muldrew_2015,Jaehyun_2024}. The lack of a hot intracluster medium in these systems means that the conventional methods of selecting galaxy clusters, such as extended X-ray emission, Sunyaev-Zel'dovich effect, or cluster red sequence will not be effective. Robust identification of protoclusters thus requires the detection of galaxy overdensities.

Different observational techniques of galaxy populations are often employed to identify protocluster regions.  These include radio galaxies \citep{Miley2008,Venemans2007,Orsi2016,Shen2021}, Lyman break galaxies \citep[LBGs: e.g.,][]{Overzier2006,Steidel1998,toshikawa18}, submillimeter galaxies  \citep[SMGs:][]{Blain2004,Dannerbauer2014,Miller2018,Pavesi2018}, H$\alpha$ emitters \citep{Hatch2011,Hayashi2012}, and Ly$\alpha$ emitters \citep[LAEs: e.g.,][]{Cowie1998,Ouchi2005,Venemans2007,lee14,Higuchi2019,shi19,Alberts_2022,Ramakrishnan_2023,Apostolovski2024,Toshikawa2024}. We refer interested readers to a review paper by \citet{Overzier_2016} for further discussions of the utility of these different types of galaxies for protocluster selection.

Ly$\alpha$ emission ($\lambda_{0} = 1215.67~\text{\AA}$) is the strongest nebular line, tracing both star formation and active galactic nuclei (AGN) activities \citep{Sobral2018,Ouchi2020}. The Ly$\alpha$ transition is a resonant line, making the interpretation more complicated. However, LAEs are typically low-dust and low-metallicity galaxies that allow Ly$\alpha$ photons to escape. At $z=2-7$, Ly$\alpha$ emission redshifts into the visible wavelengths, providing valuable means to study high-redshift galaxies \citep[e.g.,][]{malhotra02}. Existing studies suggest that LAEs represent a subset of star-forming galaxies characterized by low UV luminosity ($M_{\rm UV} \gtrsim -20)$, strong Ly$\alpha$ emission (rest-frame equivalent widths of $\gtrsim 20$~\AA), and low stellar content ($\log M_{\star}/M_\odot = 8-9$) with little-to-moderate dust extinction \citep[e.g.,][]{Ouchi2020}. These traits combined with their low galaxy bias \citep[e.g.,][]{Gawiser2007, Guaita2010, white2024} imply that they tend to be hosted by numerous, relatively low-mass dark matter halos and thus can be used as tracers of large-scale structure (LSS) of the universe. Indeed, overdensities of LAEs are linked to protoclusters and filamentary structures \citep[e.g.,][]{matsuda05,Ouchi2005,lee14, hu21, huang22,Ramakrishnan_2023}.

The One-hundred-deg$^2$ DECam Imaging in Narrowbands (ODIN) survey recently began the widest field LAE program to date. Using the Dark Energy Camera on the Blanco telescope at the Cerro Tololo Inter-American Observatory, ODIN is carrying out deep imaging of 91~deg$^2$ southern and equatorial skies with three narrow-band filters to detect LAEs at $z=2.4$, 3.1, and 4.5 (cosmic ages of 2.8, 2.1, and 1.4~Gyr, respectively). For the details of the survey design and LAE selection methods, we refer interested readers to \citet{Lee2024} and \citet{Firestone2024a}, respectively. By sampling cosmic volumes large enough to contain $\approx$600 Virgo or Coma cluster progenitors\footnote{The total comoving volume targeted by ODIN is $\approx 2\times 10^8$~cMpc$^3$. The typical angular size of each of the seven ODIN fields corresponds to 300--350~cMpc and the line-of-sight thickness is 50--75~cMpc \citep[see Table~3 in][for more detail]{Lee2024}. }, ODIN has the potential to move the field of protocluster formation forward by enabling clean and robust detection. The resultant large, uniformly selected protocluster samples cannot only facilitate meaningful statistical analyses of protocluster systems \citep{ramakrishnan24} but also elucidate the overlap between different selection methods.

This work is intended to build clear and realistic expectations for future ODIN data in finding and characterizing massive protoclusters. Additionally, we wish to forecast the formation histories of their galaxy inhabitants in order to make evolutionary connections to those in lower-redshift clusters. To track their cosmic evolution, we utilize the cosmological hydrodynamical simulation, IllustrisTNG\null. While many existing studies have taken a similar approach to study protocluster systems in general \citep{Chiang_2013,Muldrew_2015,Lovell_2017, Remus_2023}, star formation activities in protoclusters \citep{Lim_2020,Gouin_2022} and the evolution of the brightest cluster galaxies \citep{Ragone2018,Sohn_2022}, the present work is the first to evaluate the utility of LAEs and similarly low-mass star-forming galaxies as protocluster tracers across cosmic time.
We examine how the stellar population properties such as galaxy luminosity functions vary from average field to protocluster environment, and make testable predictions for future observations.

Additionally, we will focus on quantifying the impact of recent mergers in protoclusters. This is naturally of interest as galaxies in high-density protocluster regions are expected to experience higher rates of mergers than those in low-dense environments throughout their lifetime. While gas-rich mergers are thought to be linked to intense starbursts or quenching \citep[e.g.,][]{hopkins06,Pathak_2021}, which may produce extreme astronomical sources such as quasars and luminous Ly$\alpha$ nebulae, few studies thus far have been dedicated to understanding the role of mergers in the context of protocluster environment. Merger trees from cosmological simulations serve as an invaluable tool to examine the merger rates and the overall importance of mergers in protocluster observables.

This paper is organized as follows. In Section~\ref{sec:Methods},
we introduce the IllustrisTNG simulation and describe the method adopted for LAE modeling and the protocluster selection.
Section~\ref{sec:gal_pop_in_proto} covers the properties of galaxies (including LAEs) and their redshift evolution and compares them against literature measurements, while the formation histories of cluster galaxies across cosmic time are examined in Section~\ref{sec:evolutionLAEs}. Finally, in Section~\ref{sec:discussion}, we discuss the prospects of the ODIN survey in the context of our findings. Our main results are summarized in Section~ \ref{sec:conclusion}.

\section{Methods}\label{sec:Methods}

\subsection{The IllustrisTNG simulation}\label{sec:TNG}

To realistically model the formation and evolution of galaxies within the cosmological $\Lambda$CDM paradigm, we utilize the IllustrisTNG suite \citep{Nelson_2019,Pillepich_2018tng,Springel_2018,Nelson_2018tng,Naiman_2018tng,Marinacci_2018tng}, a publicly available set of magnetohydrodynamical cosmological simulations \citep{Weinberger_2017,Pillepich_2018,Vogelsberger_2014}. The IllustrisTNG simulations were run within the $\Lambda$CDM cosmology consistent with the \citet{Planck2016} with parameters ${\rm \Omega_{\Lambda,0}} = 0.6911$, ${\rm \Omega_{m,0}} = 0.3089$, ${\rm \Omega_{b,0}} = 0.0486$, ${\rm \sigma_8} =
0.8159$, $n_{\rm s} = 0.9667$ and $h = 0.6774$.
The simulations are performed with the moving-mesh code {\sc arepo} \citep{Springel_2010}. The sub-grid models include star formation, metal-line cooling, stellar feedback from supernovae Type Ia, II, and asymptotic giant branch stars, AGN feedback, and production and evolution of nine elements (H, He, C, N, O, Ne, Mg, Si, and Fe) \citep{Vogelsberger_2013,Vogelsberger_2014_Nature,Vogelsberger_2020_Nature,Barnes_2018}. 

In this work, we use the galaxy catalogs from IllustrisTNG300-1 and IllustrisTNG100-1 (hereafter TNG300 and TNG100, respectively). 
TNG300 has a side length of $302.6$~Mpc and was run with the initial conditions of (2500)$^3$ dark matter particles of mass $5.9\times10^7~\Msun$ and (2500)$^3$ gas cells of $1.1\times10^7~\Msun$. It is the largest box of the IllustrisTNG suite and contains more massive clusters that are not present in smaller-volume simulations. TNG100 offers a higher resolution at the cost of size, with a side length of $110.7$~Mpc on each side; it contains (1820)$^3$ dark matter particles of mass $7.5\times10^{6}~\Msun$, and (1820)$^3$ gas cells of $1.4\times{10^6}~\Msun$. The different particle resolutions and box sizes make them complementary to each other, which we will discuss in  Section~\ref{sec:Protocluster_Selection}.

The galaxy catalogs of TNG300 and TNG100 are available on the IllustrisTNG website and include 100 snapshots from $z=20$ to $z~=~0$ with time steps between 100--200~Myr.  Halos are identified within each snapshot through the Friends-of-Friends (FoF) algorithm  \citep{Davis_1985} on the dark matter particles with a linking length of $b=0.2$. The substructures (subhalos) are identified with the {\sc subfind} algorithm \citep{Springel_2001} run on dark matter and baryonic particles.

The mass assembly history of subhalos (and the galaxies hosted therein) is constructed through the merger tree algorithms, namely, {\sc sublink} \citep{RodriguezGomez_2015} and {\sc LhaloTrees} \citep{Springel_2005}. We opt to use the {\sc sublink} merger tree because it allows us to link the subhalos and their properties across the available snapshots. The specific methodology we used to implement the merger trees is presented in Section~\ref{sec:Protocluster_Selection}.

Finally, we note that the instantaneous star formation rates (SFRs) of the gas cells in simulated galaxies are limited by their mass resolution \citep{Donnari_2019}: the lowest measurable SFR values for TNG300 is $\log({\rm SFR}/\Msun~\yr^{-1})\sim -3$ while for TNG100 it is $\log({\rm SFR}/\Msun~\yr^{-1})\sim -4$. Galaxies with SFR values below these limits lack resolution and are given an SFR of 0. Some earlier studies allocate a random value \citep[e.g.,][$\log({\rm SFR}/\Msun~\yr^{-1})$ between $-4$ to $-5$]{Donnari_2019} to galaxies with SFR $=0$. Here, we keep the SFR value as 0 and consider them as quenched when calculating the SFR and sSFR histories for the different galaxy samples of our study.

\subsection{Modeling the LAE population in IllustrisTNG}\label{sec:model}

Two different approaches can be implemented to model LAEs in cosmological simulations. One option is through Monte Carlo-based Ly$\alpha$ radiative transfer codes \citep[see, e.g.,][]{Zheng2010,Gurung2019,Hough2020,Byrohl2023}. The other is motivated by the fact that LAEs are associated with star formation and model Ly$\alpha$ emission based on the star formation rate or UV emission from simulated galaxies \citep[see, e.g.,][]{Tilvi2009,Jose2013,Behrens2018,Weinberger_2019,Perez2021,Ravi2024}. 

Here we employ an empirical approach proposed by \citet{Dijkstra_2012} and further developed by \citet{Gronke2015,Weinberger_2019,Morales2021}. This model considers LAEs and LBGs within the same theoretical framework. The Ly$\alpha$ luminosity of the simulated galaxies is computed based on their UV luminosity combined with an observed rest-frame equivalent width (REW) probability density function (PDF). This
approach is also used for modeling the ODIN-like LAE population in the Horizon Run 5 simulation \citep{Im_2024}. We note that our methodology does not include explicit radiative transfer effect from Ly$\alpha$ photons through the large-scale intergalactic medium or diffuse gas in the protocluster volume. Ly$\alpha$ photons may undergo scattering and absorption by neutral hydrogen on large scales, which may influence our results in high-density environments \citep[see, e.g.,][]{Shimakawa_2017,PerezMartinez2023}.

For the far-ultraviolet luminosities in TNG300 and TNG100, we use results from \citet{Vogelsberger_2020}, who investigated the high-redshift galaxy population and their UV emission using three different dust attenuation models. {\sc model~a} incorporates an empirical model for dust, whereas {\sc model~b} utilizes a resolved model for dust optical depth, and {\sc model~c} employs a resolved model for dust radiative transfer.
We choose to use {\sc model~c}, the resolved dust Monte Carlo radiative transfer calculation, which takes into account stars and dust in the simulated galaxies using a modified version of the publicly available {\sc skirt} code \citep{Baes2011}. For each stellar particle, the model uses the initial mass, metallicity, and age and generates a spectral energy distribution (SED) through interpolation over a set of SED templates. Each stellar particle emits photon packets that travel through the resolved interstellar medium, randomly interacting with dust cells, before ultimately being detected by the photon detector. The model adopts a constant dust-to-metal ratio at each redshift to connect the cold gas density and metallicity to the dust density. The integrated galaxy flux generated with the {\sc skirt} is then convolved with the transmission curve of each band using the {\sc sedpy} code. The model also includes a correction for IGM absorption following \citet{Madau1995,Madau1996}.
{\sc model~c} provides the best agreement with the observed UV luminosity functions between $z=2-10$.
In this study, we classify galaxies with a UV magnitude ${M_{\rm UV}} < -18$ as star-forming galaxies.

Following \citet{Dijkstra_2012}, we adopt a PDF for the REW to determine the galaxies that  produce observable Ly$\alpha$ emission as: 
\begin{equation}
    P({\rm REW}|M_{\rm UV}) = N \exp{\left (\frac{-{\rm REW}}{{\rm REW}_{\rm c}(M_{\rm UV})}\right )}
\end{equation}
where ${\rm REW}_{\rm c}$ is a characteristic REW that depends on the UV magnitude $M_{\rm UV}$ and is expressed by the formula 
${\rm REW}_{\rm c}(M_{\rm UV}) = 23 + 7 (M_{\rm UV} + 21.9) + 6(z-4)$. The normalization $N$ is given by:
\begin{equation}
    N = \frac{1}{\rm REW_{\rm c}} \left [ \exp \left(\frac{-{\rm REW_{min}}}{\rm REW_{c}}\right ) - \exp\left(\frac{-\rm REW_{max}}{\rm REW_{c}} \right ) \right ]^{-1}
\end{equation}
where ${\rm REW_{max} = 300~\AA}$\footnote{This value corresponds to the one adopted by \citet{Dijkstra_2012} and is close to the largest REW in the sample of \citet{Ouchi_2008}. See their Section~4.2 for a further discussion.} and ${\rm REW_{min}}$ is a function of $M_{\rm UV}$:
\begin{equation}
   {\rm REW_{min}} = \left\{ \begin{array}{lcc}
             -20~\text{\AA} &     & M_{\rm UV} < -21.5; \\
             \\ 17.5~\text{\AA} &   & M_{\rm UV} > -19.0; \\
             \\ -20+6(M_{\rm UV} + 21.5)^2~\text{\AA} &    & {\rm otherwise}.
             \end{array}
   \right.
\end{equation}
The Ly$\alpha$ luminosity is then computed as follows. We first draw a random REW from the conditional probability distribution $P({\rm REW}|M_{\rm UV})$ using the $M_{\rm UV}$ of each galaxy in the TNG100 and TNG300 population. The Ly$\alpha$ luminosity is then computed as: 
\begin{equation}
    L_{\rm Ly\alpha} = c \left(\frac{\lambda_{\rm UV}}{\lambda_{\alpha}}\right)^{-\beta} \times {\rm REW} \times L_{\rm UV,\nu} 
\end{equation}
where $c$ is the speed of light and the UV spectral slope $\beta$ ($f_\lambda \propto \lambda^\beta$) is assumed to be $\beta=-1.7$. \citep{Bowman_2019}. The monochromatic UV luminosity, $L_{{\rm UV},\nu}$, and absolute UV magnitude, $M_{\rm UV}$, are measured at 1600~\AA\ and related through $M_{\rm UV}=-2.5 \log L_{\rm UV,\nu}+51.6$ \citep{Ouchi2008}. Ly$\alpha$ corresponds to frequency  $\nu_{\alpha}=2.47\times10^{15}~{\rm Hz}$ and wavelength $\lambda_{\alpha} =1215.67~\text{\AA}$.  
Having assigned the Ly$\alpha$ luminosities to all galaxies, the final step is to apply the Ly$\alpha$ luminosity and equivalent width limits of the ODIN survey, namely, $L_{{\rm Ly}\alpha} \geq 10^{42}$~erg~s$^{-1}$ and ${\rm REW} \geq 20$~\AA\ \citep{Firestone2024a}\footnote{Note that this is a simplification of the ODIN luminosity limit. In practice, ODIN utilizes a more refined selection criterion based on the narrow-band magnitude limit in each filter. This limit is sensitive to contributions from the Ly$\alpha$ emission line and the galaxy's continuum. Additionally, the REW threshold ensures the robustness of the LAE selection.}.

We repeated the above procedure five times to generate different realizations of LAEs, where all other parameters (e.g., stellar mass, star formation rate, gas mass, and UV magnitudes) remain fixed while Ly$\alpha$ luminosities change. Doing so allows us to capture the stochasticity in the Ly$\alpha$ emission for a given galaxy, rather than the intrinsic scatter across the galaxy population. Although five realizations are sufficient to illustrate the role of stochasticity and to enhance the sample size for statistical analysis, we acknowledge that this number is too small to characterize the variance introduced by the random process. 
For a more robust measurement of the stochasticity of the sample, we acknowledge that a larger number of realizations would be required. 
 Nonetheless, our approach provides a practical first-order estimate of the impact of Ly$\alpha$ stochasticity on the resulting luminosity and equivalent width distributions.

\subsection{Protocluster selection} \label{sec:Protocluster_Selection}

\begin{table}[h!]
 \caption{Properties of the most massive clusters at $z=0$ found in TNG300 (top) and TNG100 (bottom).}
    \centering
    \resizebox{0.8\columnwidth}{!}{%
     \begin{tabular}{c|S|S|S}
    \hline
   \multicolumn{4}{c}{Protoclusters - TNG300}\\
   \hline\hline
 Group ID & {M$^{z=0}_{\rm h}$} & {M$^{z=0}_{\star}$} & {SFR$^{z=0}$}  \\	[.1cm]
($z=0$) & [${10^{14}}$~\Msun] & [${10^{12}}$~\Msun] & [{\Msunyr}] \\

            \hline
    0 &  15.36 & 28.91 & 23.3\\
    1 &  13.07 & 15.95 & 31.8\\
    2 &  10.33 & 12.94 & 19.5\\
    3 &  9.00 & 12.64 & 99.3\\
	6 &  4.60 & 11.76 & 25.1\\
	4 &  8.42 & 11.24 & 44.9\\
	7 &  5.81 & 10.57 & 20.6\\
	5 &  7.34 & 10.56 & 5.0\\
	8 &  3.98 & 9.14 & 54.6\\
	10 & 6.36 & 8.70 & 14.2\\
	13 & 5.54 & 8.23 & 40.0\\
	12 & 6.55 & 8.06 & 5.3\\
	9 &  6.41 & 8.06 & 37.7\\
	11 & 5.47 & 7.57 & 7.9\\
	19 & 4.67 & 7.04 & 38.0\\
	14 & 4.37 & 6.96 & 10.1\\
	16 & 3.79 & 6.70 & 27.8\\
	15 & 4.97 & 6.68 & 24.6\\
	17 & 4.17 & 6.63 & 14.5\\
	20 & 4.31 & 6.61 & 15.0\\
	18 & 4.07 & 6.43 & 15.1\\
	22 & 3.60 & 6.26 & 25.1\\
	24 & 3.67 & 6.24 & 30.6\\
	25 & 2.20 & 6.17 & 21.4\\
	32 & 3.90 & 5.74 & 6.5\\
	29 & 2.12 & 5.70 & 18.1\\
	21 & 3.82 & 5.63 & 18.2\\
	26 & 2.78 & 5.59 & 12.0\\
	33 & 3.70 & 5.58 & 27.8\\
	23 & 3.92 & 5.48 & 16.4\\
 \hline
 \end{tabular}}
  \vspace{0.2cm}
\resizebox{0.8\columnwidth}{!}{%
 \begin{tabular}{ c|S|S|S }
 \hline
 \multicolumn{4}{c}{Protoclusters - TNG100} \\
 \hline\hline
Group ID & {M$^{z=0}_{\rm h}$} & {M$^{z=0}_{\star}$} & {SFR$^{z=0}$}  \\	[.1cm]
($z=0$) & [${10^{14}}$~\Msun] & [${10^{12}}$~\Msun] & [{\Msunyr}] \\
\hline
 0 & 3.77 & 11.04 & 98.9\\
 1 & 3.81 & 8.13 & 29.3\\
 2 & 3.38 & 7.13 & 14.2\\
 3 & 1.71 & 6.88 & 277.5\\
 5 & 2.03 & 5.04 & 28.6\\
 7 & 8.94 & 4.95 & 58.6\\
 6 & 2.09 & 4.93 & 16.8\\
 8 & 2.08 & 4.74 & 24.9\\
 4 & 2.54 & 4.55 & 42.4\\
 9 & 2.13 & 4.34 & 47.0\\
 \hline
\end{tabular}}
\tablefoot{Properties of the most massive clusters at $z=0$ found in TNG300 (top) and TNG100 (bottom). The clusters are ranked by their total stellar mass. We map the evolution of the galaxy population within a cubic volume of 60~cMpc on a side, centered on each cluster. }	\label{tab:TNG_protocluster_properties}
\end{table}

In cosmological simulations, a protocluster refers to the cores of assembling structures and all the overdensities destined to merge into a single massive, fully virialized halo by $z=0$. Observationally, the term is used more broadly. They are typically defined as extended structures composed of spatial or angular overdensities of galaxies  \citep[e.g.,][]{Steidel1998,Miller2018} although the level of overdensities and their sizes vary within the existing literature. Through this work, we intend to bridge these disparate definitions by creating realistic mock datasets that match existing observations in their global properties. 

We define a protocluster as a structure that will collapse into a galaxy cluster with a virial mass greater than $10^{14}~\Msun$ at $z=0$. Cosmological simulations show that protocluster regions observed at high redshift ($z\gtrsim 2$) exhibit a variety of evolutionary states, even when their `present-day mass' is fixed. \citet{Remus_2023} studied the evolution of the protocluster regions ranked by four different quantities -- virial mass, stellar mass, number of galaxy members, and star formation rate of the galaxy population -- and found that none of them serves as a reliable proxy for the descendant mass at $z = 0$. Thus, our $z=0$ based selection seems a prudent and robust approach that will ensure that i) the system will evolve into a massive cluster by the present time, and ii) we obtain the full range of evolutionary stages of protoclusters at high redshift.

\begin{figure}
    \centering
    \includegraphics[width=\columnwidth]{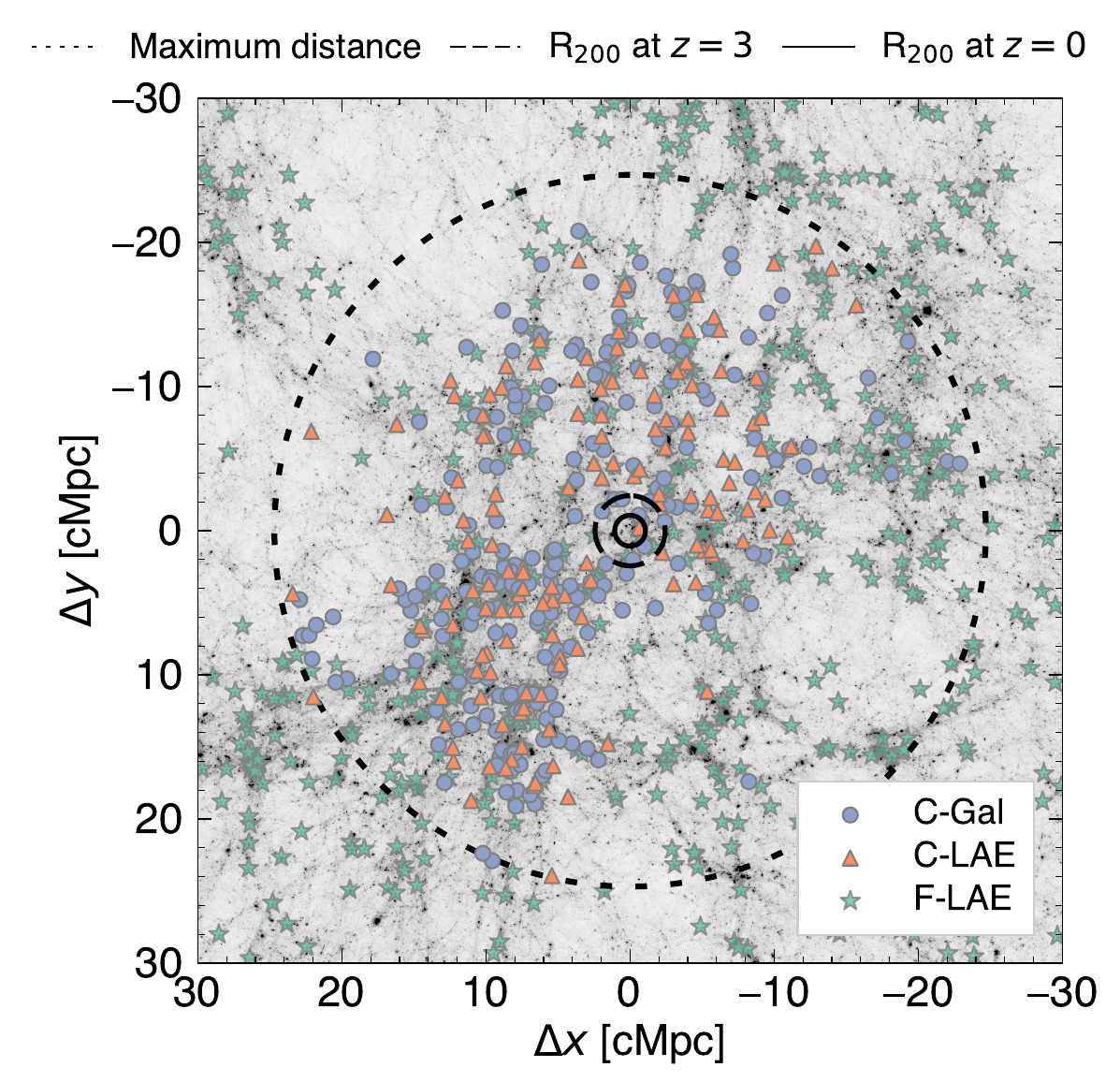}
    \caption{2D map of the galaxies at $z=3$ centered in the region of the most massive cluster of TNG300 (selected at $z=0$). Blue circles indicate the galaxies belonging to the protocluster (C-Gal). The cluster LAEs (C-LAE) are shown as orange triangles and the field LAEs (F-LAE) are shown as green stars. The outermost black short-dashed circle indicates the distance of the farthest selected galaxy belonging to the protocluster region. The inner lines indicate the $\Mhalo$ radius of the protocluster at $z=3$, and $z=0$ (long-dashed and solid-lined circles, respectively). The dark matter distribution projected across the thickness of 60 cMpc is shown in the background. 
    }
    \label{fig:z3_cluster_position}
\end{figure}

We select the 30 most massive halos in TNG300 ranked by the stellar mass at $z=0$. 
The mass of the clusters selected at $z=0$ are Virgo or Coma analogs in the range of total stellar mass M$_{\star}= (5.5-28.9)\times10^{12}~\Msun$ and M$_{\rm vir}=(3.9-15.4)\times10^{14}~\Msun$, respectively.
To complement our analysis, we also select the 10 most massive clusters of TNG100, which has a mass resolution $\sim 8$ times higher than TNG300 so can test the robustness of our findings. The mass range for the clusters in TNG100 is M$_{\star}= (4.3-11.0)\times10^{12}~\Msun$ and M$_{\rm vir}=(2.1-3.8)\times10^{14}~\Msun$. The group ID, halo mass, stellar mass, and SFR of these clusters at $z=0$ are listed in Table~\ref{tab:TNG_protocluster_properties}.

Centered around the barycenter of each cluster at $z=0$, we define a (60~cMpc)$^3$ volume in the $z=2$, 3, and 4 snapshots. The dimension of our protocluster volume matches the line-of-sight thickness of the ODIN narrow-band filters \citet{Lee2024}, to which we intend to compare our predictions in the future.
We also identify galaxies that will be cluster members at $z=0$ and trace them back in time using 
the subhalo merger trees catalog computed with {\sc sublink} \citep{RodriguezGomez_2015}. In all our analyses related to merger trees, we only track galaxies with stellar mass $10^9-10^{12}$ \Msun\ at the time of observations ($z = 2$, 3, and 4) for ease of computation. 
We select this mass range largely due to the limited resolution of TNG300. UV magnitudes can only be calculated for galaxies with a stellar mass above $10^9$~\Msun~from the radiative transfer models relying on a large number of particles. For TNG300, this limit is set to galaxies with $10^9$ \Msun~or greater as discussed in \citet{Vogelsberger_2020}.
We refer to the galaxies that are part of the massive cluster at $z=0$ as ``C-Gal.''
Similarly, the LAEs in our sample are sorted into two categories: those that became members of selected clusters at $z=0$ (C-LAEs) and those that did not (F-LAEs). The C-Gal galaxy sample includes the C-LAE sample.

In Figure~\ref{fig:z3_cluster_position}, we show the spatial distribution of cluster galaxies (C-Gals and C-LAEs) and non-cluster LAEs (F-LAEs) at $z=3$ in the most massive cluster in TNG300 (Group~0 in Table~\ref{tab:TNG_protocluster_properties}). The gray background shows the dark matter distribution at the thickness of 60~cMpc. The inner solid (long-dashed) circle represents $R_{200}$ at $z=0$ (3) while the short-dashed circle indicates the distance to the most distant cluster member galaxy. 

Besides the definitions described above, we employed additional selection criteria to create more realistic mock catalogs that can effectively mimic the observational data. 
We define field galaxies (F-Gal) as a subset of all galaxies in the full simulation box that are not within the most massive clusters selected (i.e., the 30 most massive for TNG300, 10 most massive for TNG100). 
We verify that the galaxies in the F-Gal sample are not within the most massive groups by investigating their host halo masses, $\rm M_{200}$, at $z=0$. The median host halo masses for the F-Gal galaxy sample is ${M_{200}}\sim 10^{13}$~\Msun\ at $z=0$ (see Figure~\ref{fig:TNG300_100_comp_z4}).
For the merger tree comparison, we limit the mass range of the F-Gal at each redshift to be $2\sigma$ around the median cluster stellar mass value (for each redshift, approximately $10^9 - 10^{10}\,\Msun$). This is to compare the evolution of galaxies within similar mass ranges, in and out of cluster environments. In summary, hereafter we adopt the following definitions and abbreviations to identify the galaxy populations in the simulated boxes:

\begin{itemize}
    \item F-LAE: Sample of LAEs 
    that do not become part of the selected clusters at any snapshot,\\
    \item C-LAE: Sample of LAEs that become part of a cluster at $z=0$, \\
    \item F-Gal: Field galaxies selected at $z=0$, corresponding to a random subset of galaxies outside of the clusters in the full simulation box in a refined mass range as noted above, \\
    \item C-Gal: Galaxies that become part of a cluster with $\Mstar=10^9 - 10^{12}\Msun$ at a given redshift; they include LAEs,\\
    \item PC60: Galaxies within (60~cMpc)$^{3}$ boxes centered on a cluster identified at $z=0$,\\
    \item PC15: Galaxies located in the zone of highest density within PC60 sub-boxes. For this selection, we calculate the number density within spheres of radii 7.5~cMpc centered on each galaxy. Then we select galaxies in the volume of (15~cMpc)$^{3}$ centered on the highest density. 
\end{itemize}

We also investigate the impact of mergers on the galaxy populations within clusters. For this, we use the {\sc Sublink} merger trees and identify recent major mergers that accreted at least 25\% of its mass from an outside source within the most recent snapshot ($\sim$200~Myr). 
The following abbreviations are used: 

\begin{itemize}
   \item Cluster Mergers (CM): Galaxies within the most massive clusters that experienced a major merger,\\
    \item Field Mergers (FM): Galaxies outside of most massive clusters that experienced a major merger.
\end{itemize}

\begin{figure*}
\centering
\includegraphics[width=.8\linewidth]{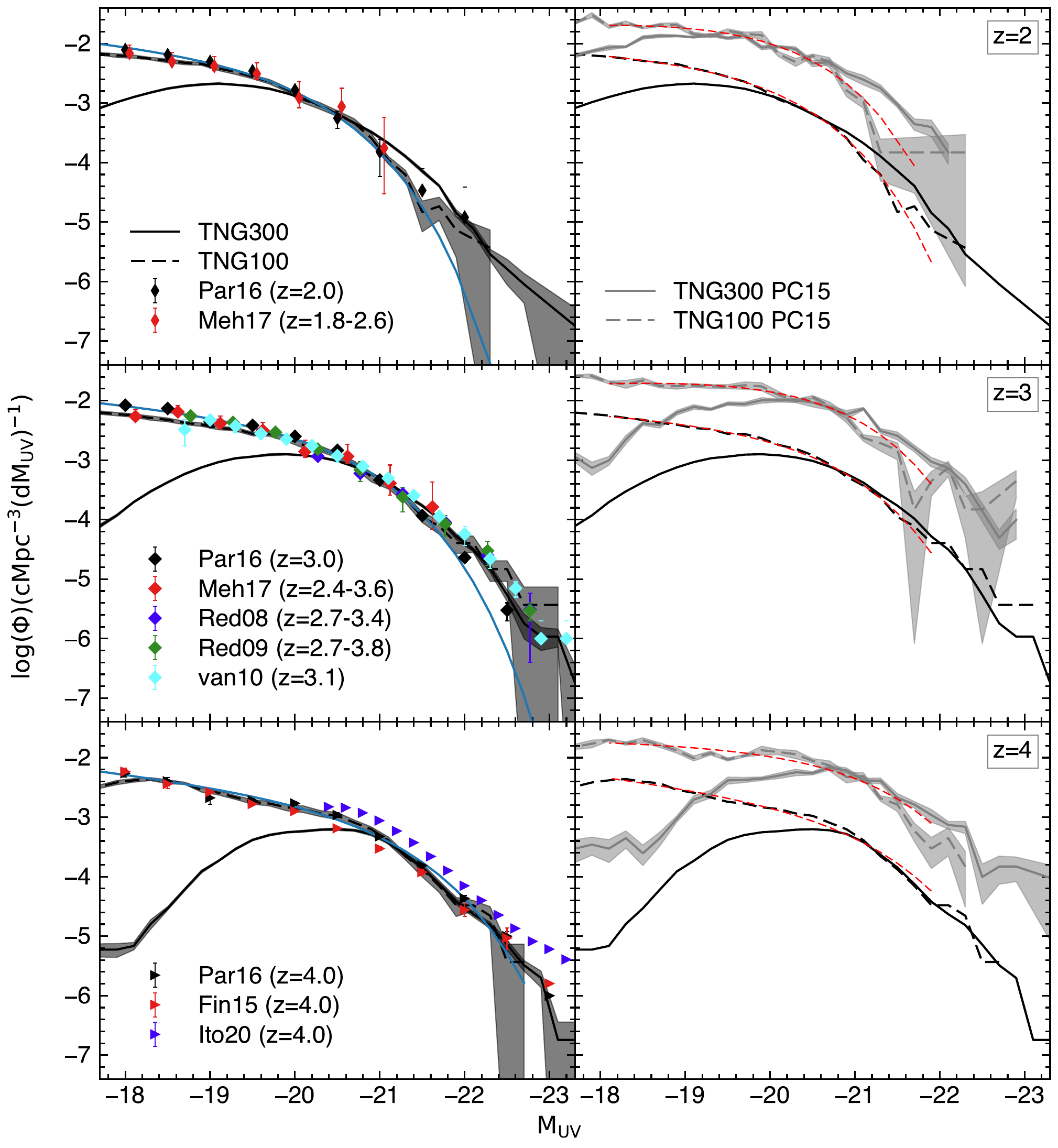}
\caption{Left: Rest-frame UV luminosity functions at $z=2, 3$, and 4 (from top to bottom, respectively) in the full volumes of TNG300 (solid) and TNG100 (dashed) in black together with the uncertainties computed via jack-knife sampling. The dropoff in TNG300 at the faint end is due to the poorer mass resolution. Various literature measurements at similar redshifts are shown, including \citet[][black symbols]{Parsa2016}, \citet[][red diamonds]{Mehta2017}, \citet[][blue and green diamonds]{Reddy2008}, \citet[][cyan diamonds]{vanderBurg2010}, \citet[][red right triangles]{Finkelstein2015}, and \citet[][blue right right triangles]{Ito_2020}. In all three panels, blue solid curves show the best-fit Schechter function of \citet{Parsa2016}.
Right: UVLFs measured within (15~cMpc)$^3$ cubic volumes centered on massive protoclusters are shown as gray lines and shades. The UVLFs for the full volumes of TNG300 and TNG100 are repeated in black. Two red dashed curves in each panel indicate the best-fit Schechter functions for the full and PC15 TNG100 measures. The faint-end slope of the protocluster UVLF is shallower than that of average fields in all cases.
}
    \label{fig:LF_UV_Protoclusters}
\end{figure*}

\section{Galaxy population in protoclusters}\label{sec:gal_pop_in_proto}

We examine the photometric and stellar population properties of TNG galaxies at $z=2$, 3, and 4. To facilitate comparisons with existing literature measurements, we analyze star-forming galaxies with and without Ly$\alpha$ emission. Particular attention is given to discerning differences in these properties along average and protocluster sightlines, aiming to establish accurate forecasts for future data, including those identified by ODIN and Rubin Observatory’s Legacy Survey of Space and Time (LSST). Leveraging merger trees, we will also study the properties of galaxies that recently underwent major mergers.

\begin{table}[h]
\caption{Best-fit Schechter parameters for TNG100-based UVLF estimates of the full volume and PC15 samples.}
\centering
\resizebox{0.5\textwidth}{!}{%
\begin{tabular}{ccccc}
\hline
Redshift  &  Data  &   $\phi^*$ & $M_{\rm UV}^\ast$ & $\alpha$\\    
 & & [$10^{-3}$~cMpc$^{-3}$] & [mag] & \\
\hline\hline
  &  \citet{Parsa2016} & $7.02\pm 0.66$ & $-19.68\pm 0.05$ & $-1.32\pm 0.03$\\
\cline{2-5}  
  & TNG100 full volume
  & $8.27\pm 0.51$ & $-19.49\pm 0.06$& $-1.03\pm 0.04$\\
$z\sim 2$& & $6.85\pm 0.11$ & $-19.68$ (fixed) & $-1.13\pm 0.01$\\
\cline{2-5}  
 &  TNG100 PC15 
 & $37.23\pm 8.30$ & $-19.73\pm 0.27$ & $-0.80\pm 0.17$\\
 &   &  $38.73\pm 2.14$& $-19.68$ (fixed) & $-0.77\pm 0.06$\\
\hline
  &  \citet{Parsa2016} & $5.32\pm 0.60$ & $-20.20\pm 0.07$ & $-1.31\pm 0.04$\\
\cline{2-5}  
  &  TNG100 full volume
  & $3.99\pm 0.41$ & $-20.38\pm 0.10$& $-1.25\pm 0.04$\\
$z\sim 3$  &   & $4.76\pm 0.11$ & $-20.20$ (fixed) & $-1.18\pm 0.02$\\
\cline{2-5}  
 &  TNG100 PC15
 & $37.51\pm 4.59$ & $-20.16\pm 0.16$ & $-0.77\pm 0.09$\\
 &   & $36.57\pm 1.49$ &  $-20.20$ (fixed) & $-0.78\pm 0.03$\\
\hline
  &  \citet{Parsa2016} & $2.06\pm 0.33$ & $-20.71\pm 0.10$ & $-1.43\pm 0.04$\\
\cline{2-5}  
  &  TNG100 full volume
  & $1.61\pm 0.54$ & $-20.91\pm 0.32$& $-1.46\pm 0.08$ \\
$z\sim 4$  &   & $1.96\pm 0.13$ & $-20.71$ (fixed) & $-1.42\pm 0.03$\\
\cline{2-5}  
 &  TNG 100 PC15
 & $12.78\pm 5.3$ & $-21.13\pm 0.50$ & $-1.18\pm 0.13$\\
 &   & $18.18\pm 1.92$ &  $-20.71$ (fixed) & $-1.06\pm 0.06$\\
\hline
\end{tabular}}
\tablefoot{For ease of comparison, we list the observational values from \citet{Parsa2016}. For each TNG100-based UVLF, we tabulate two sets of fitting parameters; one while keeping all parameters free during fitting, and another while fixing $M_{\rm UV}^\ast$ from \citet{Parsa2016}.}
\label{tab:UVLF}
\end{table}

\subsection{Luminosity functions}\label{subsec:LF_field}

In the left panels of Figure~\ref{fig:LF_UV_Protoclusters}, we show the rest-frame UV luminosity functions (UVLFs) measured from the full TNG100 and TNG300 simulations. The error bars, indicated by gray strips, are calculated via jack-knife resampling by dividing the volume into eight sections.  Color symbols show the literature measurements taken from \citet{Reddy2008, Reddy2009, vanderBurg2010, Finkelstein2015, Parsa2016, Mehta2017}. Analogous to the result reported by \citet{Vogelsberger_2020}, our TNG100 measurements reproduce these observations reasonably well over $M_{\rm UV} = -(18 - 23)$. Our TNG300 results generally agree with TNG100 but diverge at $M_{\rm UV} \gtrsim -20$. This is an expected consequence of the poorer mass resolution of TNG300, which results in fewer low-mass halos therein. 

We fit the simulation data to a \citet{Schechter76} function, using the $M_{\rm UV}$ in the range $-18 < M_{UV} < -22$ and list the best-fit parameters in Table~\ref{tab:UVLF}.
The faint-end slope $\alpha$ is generally found to be shallower than those in the literature. When we fix $M_{\rm UV}^\ast$ to the values tabulated in \citet{Parsa2016}, the best-fit $\alpha$ value changes only slightly.
Since the \citet{Parsa2016} measurements cover all three redshift ranges and are not discrepant with other measures in the literature, their best-fit Schechter functions are also illustrated in the left panels of Figure~\ref{fig:LF_UV_Protoclusters}. The shallower $\alpha$ value is consistent with the fact that most data points (especially at $z=3$) lie above the TNG curves at a $\approx$20\% level.

We also compute the Ly$\alpha$ luminosity functions (LALFs), for which we only use galaxies selected as LAEs as described in Section~\ref{sec:model}. We include LAEs from all five independent realizations in which Ly$\alpha$ luminosities are assigned to galaxies according to a conditional probability density. Following \citet{Dijkstra_2012}, we include a normalization factor $F$. As the ratio of
the observed to predicted number density of LAEs, $F$ may be considered as a fudge factor reflecting the uncertainties associated with the complexities of Ly$\alpha$ radiative transfer, the internal structure of dust and gas, and the star formation histories (and duty cycle) of the galaxies we sample. To compare our findings with observations  \citep{Dawson2007,Ouchi2008,Zheng2013,Konno2016,Sobral2018}, we apply the minimum equivalent width of 60~\AA\ at $z=2$ \citep{Konno2016}, 64~\AA\ at $z=3$ \citep{Ouchi2008}, and 9~\AA\ at $z=4$ \citep{Zheng2013}, respectively, and find $F\approx 0.51$ in agreement with \citet{Dijkstra_2012}. 

\begin{figure*}[h]
\centering
\includegraphics[width=.8\linewidth]{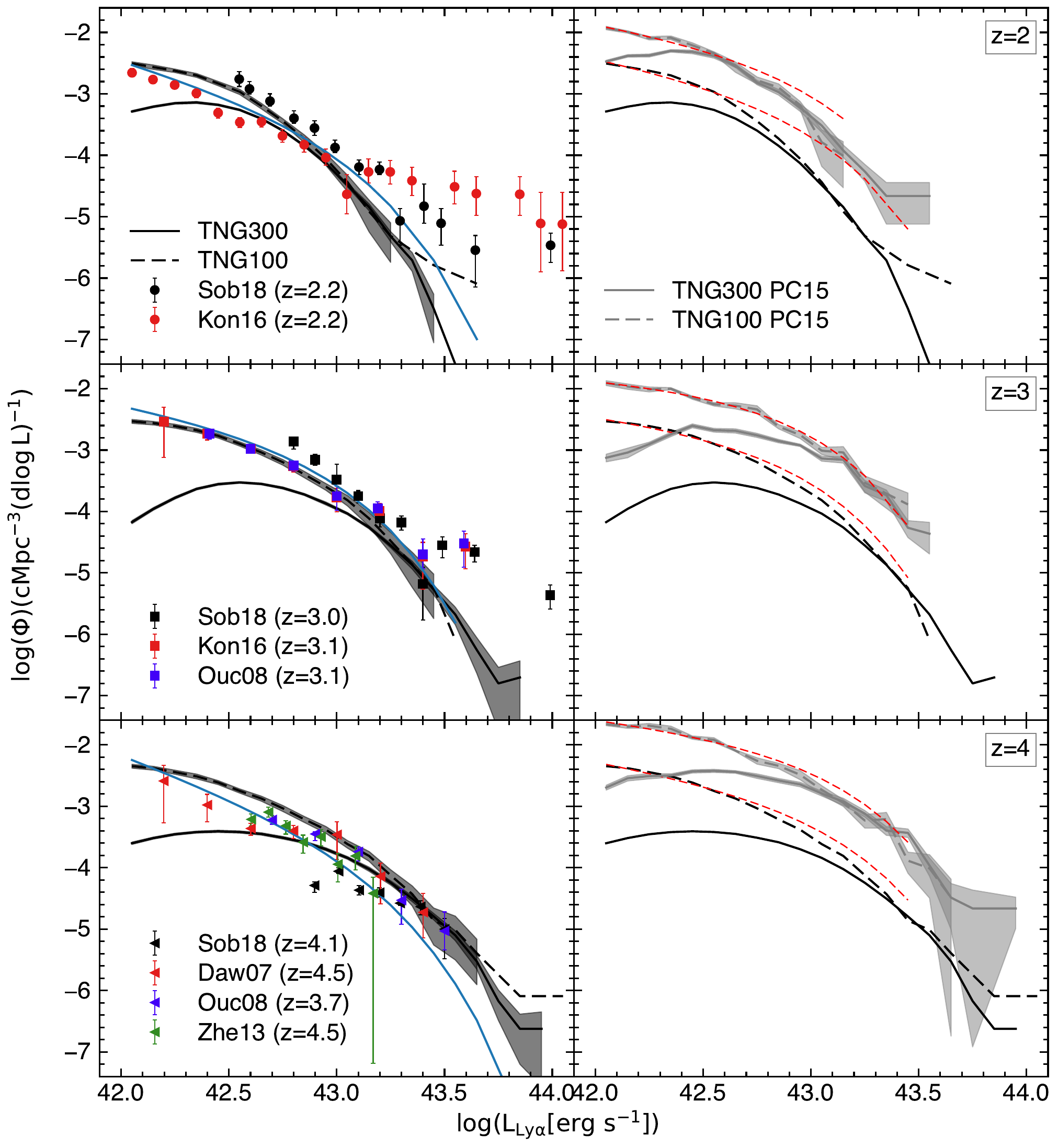}
\caption{Left:
Ly$\alpha$ luminosity functions (LALFs) at $z=2, 3$, and 4 (from top to bottom, respectively) in the full volumes of TNG300 (solid) and TNG100 (dashed) in black together with the uncertainties computed via jack-knife sampling. Like the UVLF case, the TNG300-based measures are underestimated at the faint end due to the poorer mass resolution. 
Literature measurements include \citet[][blue symbols]{Sobral2018}, \citet[][orange circles and squares]{Konno2016}, \citet[][green squares and triangles]{Ouchi2008},  \citet[][orange triangles]{Dawson2007}, \citet[][red triangles]{Zheng2013}. In all three panels, blue solid curves show the best-fit Schechter function of \citet{Sobral2018}.
Right: LALFs measured within (15~cMpc)$^3$ cubic volumes centered on massive protoclusters are shown as gray lines and shades (PC15 boxes). The field LALFs are repeated in black. Two red dashed curves in each panel indicate the best-fit Schechter functions for the TNG100 measures. The lower red-dashed curve is for the field regions and the higher red-dashed curve is for the protoclusters. 
}
    \label{fig:LF_LAE_Protoclusters}
\end{figure*}

The left panels of Figure~\ref{fig:LF_LAE_Protoclusters} show the LALFs adjusted for the $F$ factor. 
The agreement with the data is poorer than the UVLF case, which in part reflects the disagreement between existing measurements. For example, the shapes of the LALF from \citet{Sobral2018} and \citet{Konno2016} are notably dissimilar. 
The apparent disagreement between the real data and simulations at the bright end ($L_{{\rm Ly}\alpha}\gtrsim 10^{43}$~erg~s$^{-1}$) is unlikely to be real as photometrically selected LAE samples are often contaminated by low-luminosity AGN with broad Ly$\alpha$ emission \citep[e.g.,][]{Konno2016}. The excess at the bright end renders the shape of the LALF to resemble a `double power-law' rather than a Schechter form.  
To our knowledge, no semianalytical models or hydrodynamic simulations have been able to reproduce the observed bright-end excess of the luminosity functions due to AGN. When these complications are considered, the TNG100 predictions are in reasonable agreement with existing observations.
Once again, the TNG300 predictions tend to underestimate the number density at the faint end compared to both TNG100 and observational data.

Given relatively large disparities among existing LALF measurements, we do not present the best-fit Schechter parameters in a table form. 
The best-fit Schechter functions obtained for our TNG100 LALFs are illustrated as red dashed lines in the right panels of Figure~\ref{fig:LF_LAE_Protoclusters}, while those from \citet[][]{Sobral2018} are shown as blue solid curves in the left panels. When we fix $L_{{\rm Ly}\alpha}^\ast$ to the values given in \citet[][]{Sobral2018}, i.e., $10^{42.82}$, $10^{42.77}$, and $10^{42.93}$~${\rm erg\,s}^{-1}$  at $z=2.2, 3.1, 3.9$, respectively (see their Table 6), the faint-end slopes of TNG100 galaxies are $-1.67\pm 0.07$, $-1.41 \pm 0.04$, $-1.59 \pm0.04$,   broadly consistent with the typically quoted range of $\alpha$ values, $-(1.6-1.8)$, adopted for the observational data \citep[e.g.,][]{Ouchi2008,Sobral2018}.

\label{sec:UVLF_for_fornax_PC}
\begin{figure}
    \centering
    \includegraphics[width=\columnwidth]{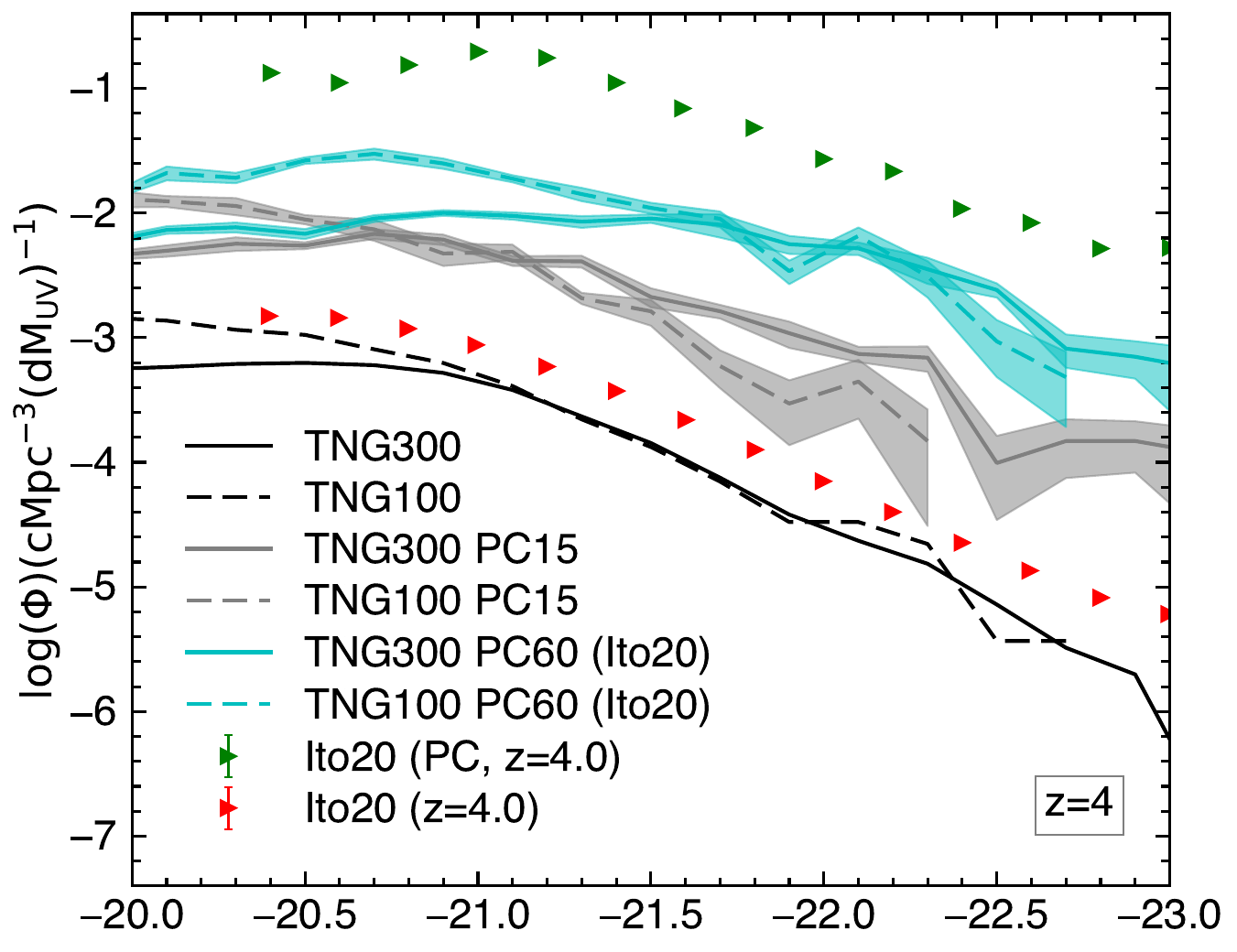}
    \includegraphics[width=\columnwidth]{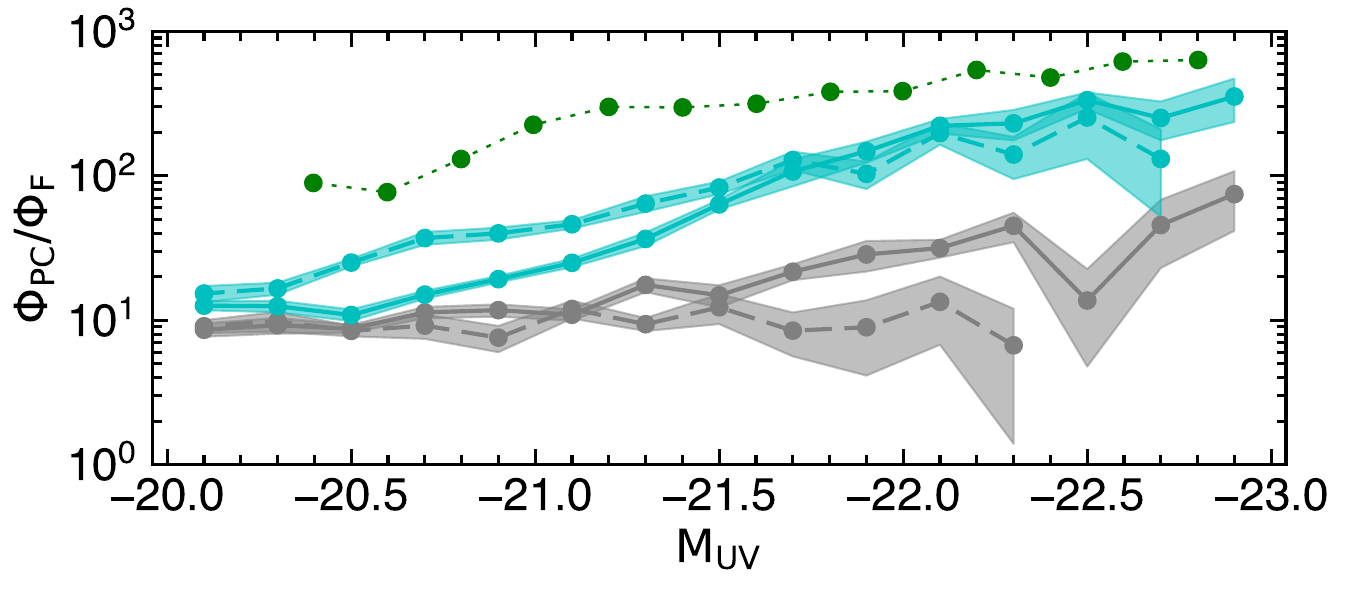}
    \caption{
    Top: UVLFs of galaxies at $z=4$ in the field (black) and protocluster (gray) volumes. The observational measures from GOLDRUSH \citep{Ito_2020} are shown as green and red triangles. The revised measurements when protoclusters are selected similar to their approach are shown in cyan lines and shades. In protocluster environment, the overall number density is much greater.
    Bottom: the ratio of the protocluster-to-field UVLF indicates the excess of UV-bright galaxies. The trend is even more pronounced when protoclusters are defined as regions of enhanced surface density in smaller areas (corresponding to a cylindrical volume of 7.5~cMpc in diameter and depth). 
    }
    \label{fig:UVLF_for_fornax_PC}
\end{figure}

\subsection{Protocluster luminosity functions}\label{subsec:pclf}

To develop expectations for LFs in protocluster volumes, we repeated the measurements using PC boxes. 
To roughly match the effective radii (enclosing about half of the total mass) of $z\sim 3$ protoclusters, which range from 5--8~cMpc \citep{Chiang_2013},
we consider a (15~cMpc)$^3$ sub-volume centered on each of our 30 protoclusters. 
The results for the UVLFs and LALFs are shown as gray lines and shaded regions in the right panels of Figures~\ref{fig:LF_UV_Protoclusters} and \ref{fig:LF_LAE_Protoclusters}, respectively. The full simulation measurements -- which can be considered as the `blank average field' measures --  are repeated in black solid and dashed lines. 

As expected, the overall normalization of the LF is higher in protocluster regions by approximately an order of magnitude. The high normalization reflects the level of three-dimensional galaxy overdensity within a volume encompassing about half of the total $z=0$ mass. Observational measurements based on the LAE- or LBG distribution would be lower \citep[e.g.,][]{lee14,toshikawa18,Ito_2020} due to unassociated interlopers in their sightlines. For example, ODIN uses filters with $\Delta \lambda \sim 70-100$~\AA\ corresponding to the line-of-sight thickness of $\sim$60--70~cMpc. We return to this topic in Section~\ref{sec:discussion}. 

We report a clear difference in the shapes of the average- and protocluster UVLFs. The best-fit Schechter parameters from TNG100 galaxies are summarized in Table~\ref{tab:UVLF}.  The difference in the faint-end slope is most clearly visible at $z=2$ and 3 (dashed lines in the top and middle right panel of Figure~\ref{fig:LF_UV_Protoclusters}), where $\alpha$ is consistently shallower in protocluster volumes by $\Delta \alpha \approx 0.3-0.4$ compared to the full TNG100 volume. 
Fixing the characteristic luminosity to that of \citet{Parsa2016}, the slope does not change dramatically. 
A similar trend persists for the LALFs. The faint-end slopes are  $\alpha_{\rm PC}$= $-0.80 \pm 0.17$, $-0.77\pm 0.09$, and $-1.18 \pm 0.13$ for protoclusters, compared to $-1.03 \pm 0.04$, $-1.25 \pm 0.04$, and $-1.46 \pm 0.08$ for the full TNG100 box at $z=2$, 3, and 4, respectively. 

We also observe an excess number of galaxies on the bright end of the UVLF\null. In the top panel of Figure~\ref{fig:UVLF_for_fornax_PC}, we compare the $z=4$ UVLFs within the (15~cMpc)$^3$ protocluster volumes (gray lines) with the full-volume UVLFs (in black). The observational measurements overlaid in the figure include the protocluster-  and average LF from \citet{Ito_2020} as green and red triangles, respectively. In the bottom panel, we show the ratio of protocluster-to-field number densities in each magnitude bin. At $M_{\rm UV} \simeq -20.0$, the overall enhancement is $\approx$9. Moreover, the enhancement increases toward higher luminosity up to $\approx$25 at $M_{\rm UV} \approx -22$. The higher enhancement seen in TNG300 is explained by the fact that TNG300 contains a larger number of massive structures than TNG100. In terms of the present-day mass, the TNG300-selected structures range in $M_h^{z=0} = (2-15)\times 10^{14}M_\odot$ whereas the TNG100 structures range in $M_h^{z=0}=(2-4)\times 10^{14}M_\odot$ as listed in Table~\ref{tab:TNG_protocluster_properties}. For the UVLFs at $z=2$ and 3, we observe similar trends. 

The excess of UV-luminous galaxies found in TNG simulations signals enhanced star formation in massive protoclusters, in broad agreement with the result of \citet{Ito_2020}. While our analyses focus on the systems that will evolve to massive clusters at $z=0$, the protoclusters used by \citet{Ito_2020} are selected as LBG overdensities \citep{toshikawa18}. Several key differences complicate a direct comparison of our results with theirs. First, the redshift selection function of LBGs is much wider ($\Delta z \approx 1$) than that of LAEs ($\Delta z \approx 0.06$ for ODIN), and thus the same protocluster would appear as a less significant overdensity with a smaller angular extent in the LBG-based selection \citep[e.g., compare Figure~7 and Figure~5 in][which show the density maps from the LAE and LBG-based selection, respectively]{Ramakrishnan_2023,toshikawa18}. Even the largest TNG simulation does not cover sufficient line-of-sight depth to closely match their redshift selection function. Secondly, our measurements are for the central (15~cMpc)$^3$ volume of each protocluster, which not only contains the highly overdense protocluster core but also lower-density regions between halos (see Figure~\ref{fig:z3_cluster_position}). In contrast, \citet{Ito_2020} selected each protocluster as a region of high overdensity within a cylindrical volume of 7.5~cMpc in both diameter and depth. 

We recompute the protocluster UVLF using a procedure similar to that described in  \cite{Ito_2020} as follows. 
First, we split each of the PC boxes into eight 7.5~cMpc-thick `slices' along the $z-$axis. Each slice is then projected onto the x-y plane to produce a 2D sky image. 
We place circular apertures of 7.5~cMpc in diameter, centering them in 2.08~cMpc grid cells, and measure the mean ($\mu$) and the standard deviation ($\sigma$) of the LAE number counts. Each overdensity is defined as a region that rises with a significance of at least $4\sigma$, similar to the protocluster sample selected by \citet{toshikawa18}. The new UVLF is computed using the galaxies that are inside these protocluster boundaries.

that gave an new way.

The revised UVLF measures are shown in  Figure~\ref{fig:UVLF_for_fornax_PC} as cyan lines and shaded regions. The overall normalization relative to the field is even more pronounced than previously. We do not reproduce the extremely high overdensity values ($\approx 100-400$) reported by \citet{Ito_2020}, which may be in part because they tend to pick up more massive protoclusters than those in our sample. Given a much larger redshift selection function of LBG-selected galaxies, fore- and background contaminants in their sample could also contribute to this discrepancy. Furthermore, our approach using TNG volumes is inadequate to simulate the wide redshift selection of the LBG-based protocluster search ($\Delta z \approx 1$), and alternative methods such as light cones are needed.

\begin{figure}
\centering
    \includegraphics[width=\linewidth]{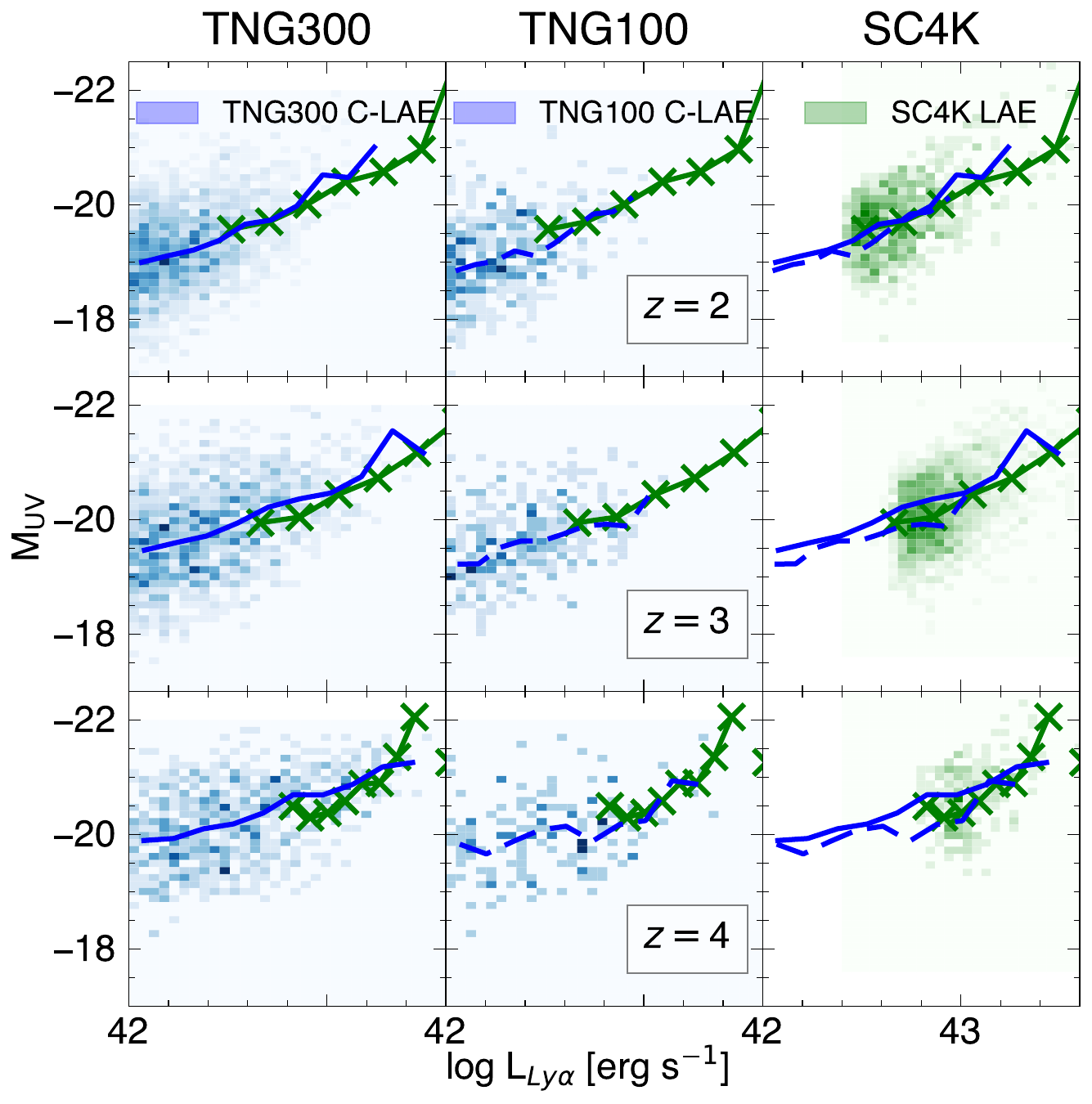}
  \caption{Comparison of Ly$\alpha$ Luminosity vs. $\rm{M}_{UV}$ for the C-LAEs in TNG300 and TNG100 at redshifts $z=2$, 3, and 4 (from top to bottom). We compare our results with the observations from the \citet{Sobral_2018a} SC4K LAE catalog in green. The redshift bins compared to each redshift follow those in \citet{Sobral_2018a} as $z = 2.22\pm 0.02$ compared to $z = 2$ in TNG300 (top), $z = 3.1 \pm 0.4$ (middle), and $z = 3.9 \pm 0.3$ (bottom). The lines in each row compare the median LAEs at each luminosity (solid blue line for TNG300, dashed blue line for TNG100) to the observed SC4K LAE (green lines with cross hatch).
  }
  \label{fig:MUV_comp}
\end{figure}

The excess on the bright end of the UVLF is also greater in the revised measurement where, for both TNG simulations, the enhancement increases monotonically toward higher UV luminosities. 
The protocluster-to-field ratio of integrated UVLF at $M_{UV}=-(23-20)$ is  $\approx$24 and 32 for TNG300 and TNG100, respectively. The ratio based on the \citet{Ito_2020}-like approach in the same magnitude range gives 
$\approx$180. In Section~\ref{subsec:merger}, we show that higher merger rates in protocluster environments are in part responsible for enhanced star formation activities (i.e., their excess of UV-bright galaxies).

\subsection{The properties of galaxies in protoclusters versus field}\label{subsec:global_properties_CLAE}

While protoclusters are rich in star-forming galaxies such as LAEs and LBGs, they are expected to evolve into a region largely populated by massive quenched galaxies by $z=0$.
In this section, we study the properties of the population of galaxies associated with massive clusters at $z=0$ at $z=2,3$ and 4. In particular, we focus on the LAE population, namely the C-LAE galaxies, and compare them with galaxies in the field.

\begin{figure*}[h!]
    \centering
\includegraphics[width=0.6\textwidth]{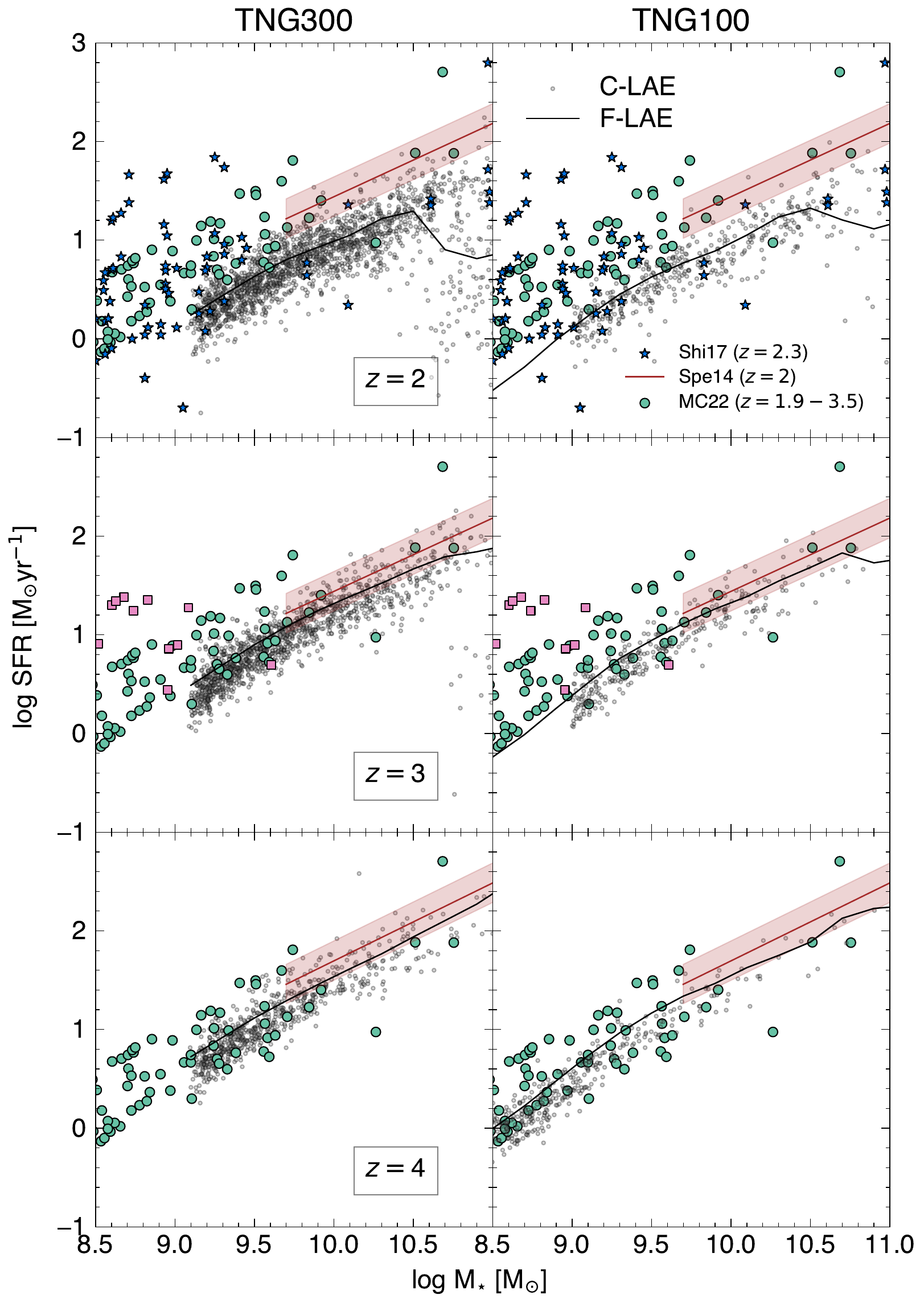}
  \caption{SFR vs. M$_{\star}$ for TNG300 (left panels) and TNG100 (right) at $z=2, 3,$ and 4 (from top to bottom) are shown. The black line illustrates the median scaling law for F-LAEs while the small gray dots show the C-LAEs.  
Existing measurements from \citet[][$z=2.3$: blue stars]{Shimakawa_2017}, \citet[][$z=1.9-3.5$: green circles]{McCarron_2022}, and \citet[][$z=2.65$: pink squares]{Pucha_2022} are in the appropriate redshift panels. The corresponding star-forming main sequence and its $1\sigma$ scatter from \citet{Speagle_2014} are indicated by salmon lines and swaths.
  }
  \label{fig:obscomp_hist_z234}
\end{figure*}

We first investigate the $M_{\rm UV}$ and $L_{{\rm Ly}\alpha}$ of the C-LAEs. 
Both UV and Ly$\alpha$ luminosities trace recent star formation in galaxies and are expected to be correlated. Observations show that LAEs tend to be less UV-bright and more compact than other galaxy populations \citep[see, e.g.,][]{Malhotra2012}. 
In Figure~\ref{fig:MUV_comp}, we present the $M_{\rm UV}$ versus $L_{{\rm Ly}\alpha}$ relation for the C-LAEs at $z=2,3$, and 4. The measurements for the SC4K LAEs from \citet{Sobral_2018a} are shown in the right panels. The median values of $M_{\rm UV}$ in each $\log L_{{\rm Ly}\alpha}$ bin are shown for TNG300 (blue solid) and TNG100 (blue dashed) together with the \citet{Sobral_2018a} data (green solid line). Our results show good agreement indicating that more UV-luminous LAEs also tend to have higher Ly$\alpha$ luminosities. The correlation between Ly$\alpha$ and $M_{\rm UV}$ is expected, first because our probabilistic LAE modeling is based on the $M_{\rm UV}$ of the simulated galaxies (see Section~\ref{sec:model}), and second because LAEs are star-forming galaxies harboring massive young stars responsible for producing the UV continuum. Nevertheless, it is reassuring that our formalism, calibrated with observational data from more than a decade ago \citep[see][]{Dijkstra_2012}, produces results that fully agree with more recent data. If one were to perform this experiment with LBGs, this correlation would not be observed since LAEs are preselected to have lower dust attenuation. The large spread of $\sim 1-3$~dex in UV luminosities at a fixed Ly$\alpha$ luminosity is also qualitatively consistent with observational studies \citep[see discussion in][]{Matthee2017,Sobral2018}.

In Figure~\ref{fig:obscomp_hist_z234}, we show the SFR and $M_{\star}$ values of C-LAEs at $z=2, 3$, and 4 (as gray dots) and compare them with the F-LAEs sample. For clarity, we display the median scaling relation for F-LAEs using black lines instead of individual data points.
We do not find any significant difference in the SFR-$M_{\star}$ scaling between C- and F-LAEs at any redshift. 
While observational data for comparison is rather limited, we show existing measurements from \citet[][$z=2.3$: blue stars]{Shimakawa_2017}, \citet[][$z=1.9-3.5$: green circles]{McCarron_2022}, and \citet[][$z=2.65$: pink squares]{Pucha_2022} together with the star-forming main sequence (SFMS: salmon lines and swaths) from \citet{Speagle_2014} at the appropriate redshift panels. In all cases, the \citet{chabrier03} stellar initial mass function is assumed. At $z=3$ and 4, the agreement between TNG galaxies and real data is reasonable. However, at $z=2$, both C-LAEs and F-LAEs underpredict SFR values at a fixed stellar mass. Additionally, at the high-mass end ($\log M_\star/M_\odot \gtrsim 10.5$), the SFR shows a dip, suggesting that some of the massive galaxies may begin to quench at $z=2$,  
perhaps due to AGN feedback \citep[see][for a further discussion on quenched galaxies in IllustrisTNG]{Donnari2021}. It is also possible that other quenching mechanisms may also play a role, such as the growth of stellar bulges \citep{Schreiber_2015}.

All in all, the properties of TNG galaxies strongly suggest that the star-forming main sequence is shared by field and protocluster galaxies alike. However, our finding does not necessarily preclude the possibility of enhanced star formation in protoclusters \citep[see, e.g.,][]{dey16,lemaux22,wells22} nor is it inconsistent with the excess of UV-luminous galaxies in a dense environment discussed in Section~\ref{subsec:pclf}. Protocluster galaxies may simply slide along the star-forming main sequence toward the high SFR and high $M_\star$ end while steadily growing at higher rates than other galaxies \citep{shi19}. This means that perhaps a more robust way to probe the environmental effects on galaxy formation may be to focus on the overall distribution of SFR or luminosity \citep[as described in Section~\ref{subsec:pclf} and in][]{Ito_2020} rather than the SFR-$M_\star$ relation. The large and contiguous area coverage of ODIN will facilitate clear and direct comparisons, especially after the arrival of LSST and the Nancy Grace Roman telescope.

\subsection{Galaxy mergers in protoclusters}\label{subsec:merger} 

In Section~\ref{subsec:global_properties_CLAE} and Figure~\ref{fig:obscomp_hist_z234}, we find that protocluster and field galaxies in the TNG simulations are found in similar locations on the SFR-$M_\star$ plane, forming a common star-forming main sequence \citep[SFMS; e.g.,][]{Speagle_2014}. The implication is that a relatively small fraction of galaxies even in dense protocluster environments are dominated by bursty star formation, which would place them above the SFMS. Possible drivers of starbursts include galaxy mergers and close galaxy-galaxy interactions, which lead to the loss of angular momentum in cold gas \citep[see, e.g.,][]{Kennicutt1987,Mihos1996,Elbaz2003}. As galaxy mergers are expected to be more frequent in dense environments, it may contribute to the increased star formation activity.

\begin{figure}
    \includegraphics[width=\columnwidth]{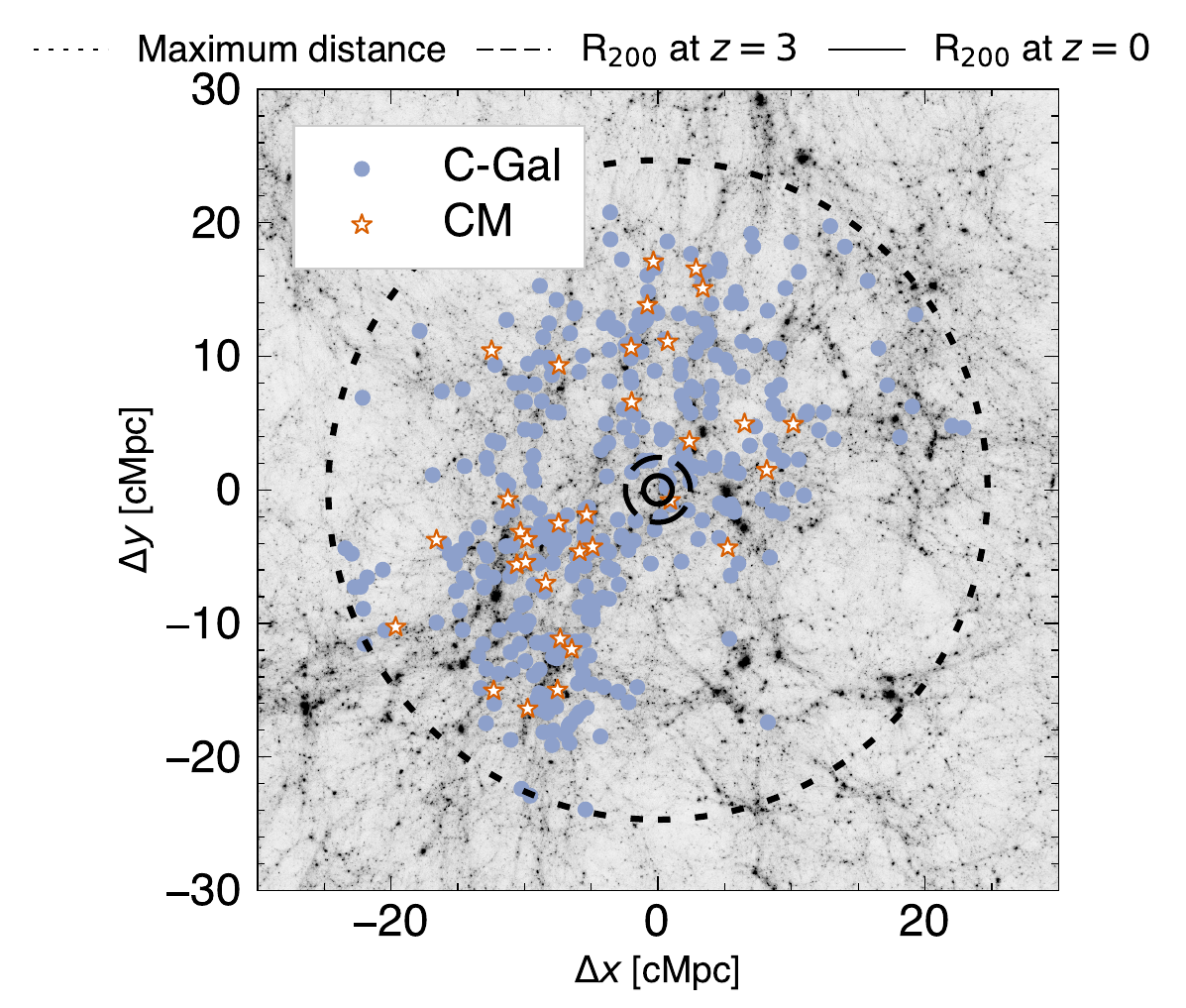}
  \caption{Spatial distribution of galaxies at $z=3$ around the most massive cluster in TNG300 (Group 0, see Table~\ref{tab:TNG_protocluster_properties}, same as  Figure~\ref{fig:z3_cluster_position}). 
  Orange stars show the cluster member (CM) galaxies that underwent a recent major merger -- defined as a mass ratio of 1:$n$ ($n\leq 4$) -- in the last 200~Myr while blue circles indicate those that did not (C-Gal). The dark matter distribution at the thickness of 60~cMpc is shown in the background. The black circles are the same shown in Figure \ref{fig:z3_cluster_position}, indicating the farthest galaxy position, $R_{200}$ at $z=3$, and $R_{200}$ at $z=0$.
  }  
  \label{fig:recent_merger_position}
\end{figure}

Here we use the subhalo merger trees to investigate the contribution of major mergers to the star formation rate in galaxies. A galaxy is considered to have experienced a recent merger if the host (sub)halo merged with another(sub)halo with a mass ratio of 1:$n$ ($n\leq 4$) within the past $\sim$200~Myr. Samples of galaxies with recent mergers are identified at the $z=2$, 3, and 4 snapshots from both TNG300 and TNG100 and then divided into two groups: those in clusters (CM for cluster merger) and field (FM for field merger). 
Figure~\ref{fig:recent_merger_position} illustrates the spatial distribution of the CM and C-Gal samples around one of the protocluster regions at $z=3$ of TNG300. We find that the angular distribution of the CM sample generally traces that of the C-Gal sample. 
The figure also shows that the occurrences of major mergers are relatively rare even in highly dense protocluster environments. 

\begin{figure*}[!htbp]
    \centering
    \includegraphics[width=.85\linewidth]{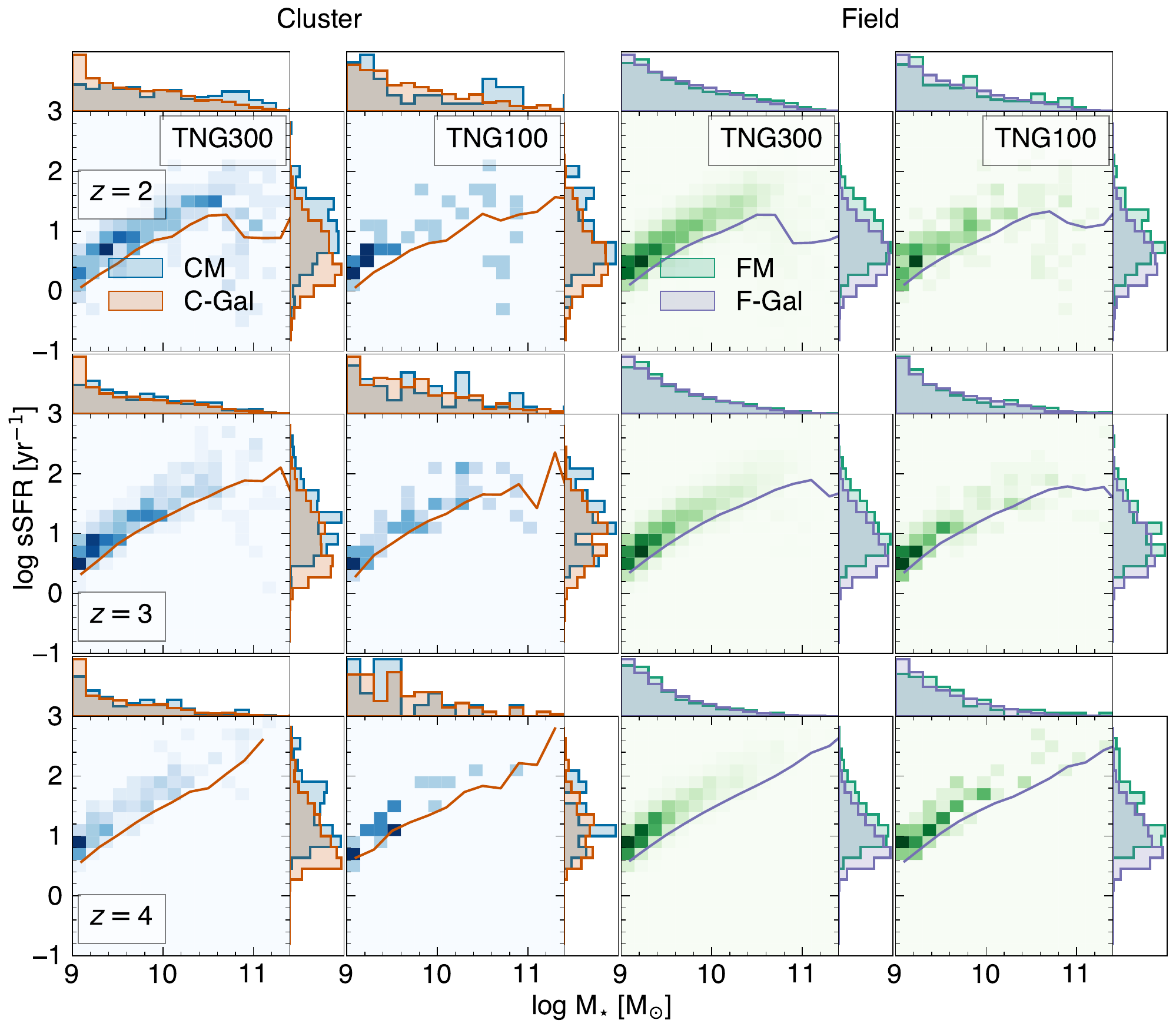}
  \caption{
  Main panels:
  $\SFR$ vs. $\Mstar$ for galaxies that experienced a major merger in the last $\sim$200~Myr at the time of observations at $z=2$, 3, and 4. 
  First and second columns present the distribution for CM (blue shades) and C-Gal (orange lines) for TNG300 and TNG100, respectively. The third and fourth columns display the distribution for FM (green shades) and F-Gal (light violet lines) for TNG300 and TNG100, respectively.
  Top and right side panels:
  Marginalized histograms for the stellar mass and SFR of the different galaxy samples at $z=2, 3,$ and 4 for TNG300 and TNG100. 
  The overall distributions of SFR and $\Mstar$ for the CM and FM galaxies are markedly different from those of C-Gal and F-Gal. In particular, CM galaxies show a significant excess on the high-mass end in their mass distribution with respect to the C-Gal population. Furthermore, among the CM, FM, C-Gal and F-Gal groups, we find that the median SFR and $\Mstar$ values are the highest for CM galaxies.
  }
  \label{fig:merger_comp}
\end{figure*}

In the main panels of Figure~\ref{fig:merger_comp}, we show the locations of galaxies on the SFR-$\Mstar$ plane. The CM galaxies are shown in the first and second column color-coded by density, for TNG300 and TNG100, respectively, as blue 2D histograms in the main panel. The orange lines show the distribution of C-Gal (i.e., cluster galaxies that have not experienced a recent merger for comparison). Similarly, the FM galaxies are represented in the third and fourth columns in green where F-Gal (field galaxies without recent mergers) are shown as purple lines. 

In all cases, we observe an enhancement of SFR at the $\approx$0.2~dex (60\%) level at a fixed stellar mass regardless of the environment, suggesting that mergers in any environment uniformly boost star formation activity. This is consistent with observations from $0<z<1.2$ \citep{Hwang_2011}.

The subpanels of Figure~\ref{fig:merger_comp} show that the overall distributions of SFR and $\Mstar$ for the CM and FM galaxies are markedly different from those of C-Gal and F-Gal. All histograms are plotted on a linear scale and normalized so that the area under it is unity. 
While a subset of galaxies in both field- and cluster environments underwent recent mergers, the top subpanels show that only CM galaxies show a significant excess at the high-mass end in their stellar mass distribution, with CM galaxies being on average $\sim0.1-0.4$~dex more massive than the other samples.
Similarly, the right subpanels suggest that CM galaxies tend to have a more pronounced tail of high-SFR galaxies than their FM counterparts. Among the CM, FM, C-Gal, and F-Gal groups, the median stellar mass and SFR values are the highest for CM galaxies.
The enhancement in SFR for CM galaxies is attributed to their higher stellar masses, rather than by an intrinsic difference in the specific star formation rate, which remains similar across the CM and FM populations for the three redshifts and in both simulations.

The different distributions of SFR and $\Mstar$ for CM galaxies are in line with the different UVLFs and LALFs in protocluster volumes (Figures~\ref{fig:LF_UV_Protoclusters}, \ref{fig:LF_LAE_Protoclusters}, and \ref{fig:UVLF_for_fornax_PC}). 
Our findings are in qualitative agreement with the expectation that deeper potential wells and higher merger rates of halos in dense protocluster environments favor the formation of massive galaxies while suppressing that of low-mass galaxies compared to the average field.

\section{Formation histories of protocluster galaxies}\label{sec:evolutionLAEs}

\begin{figure*}
    \centering
    \includegraphics[width=.45\textwidth]{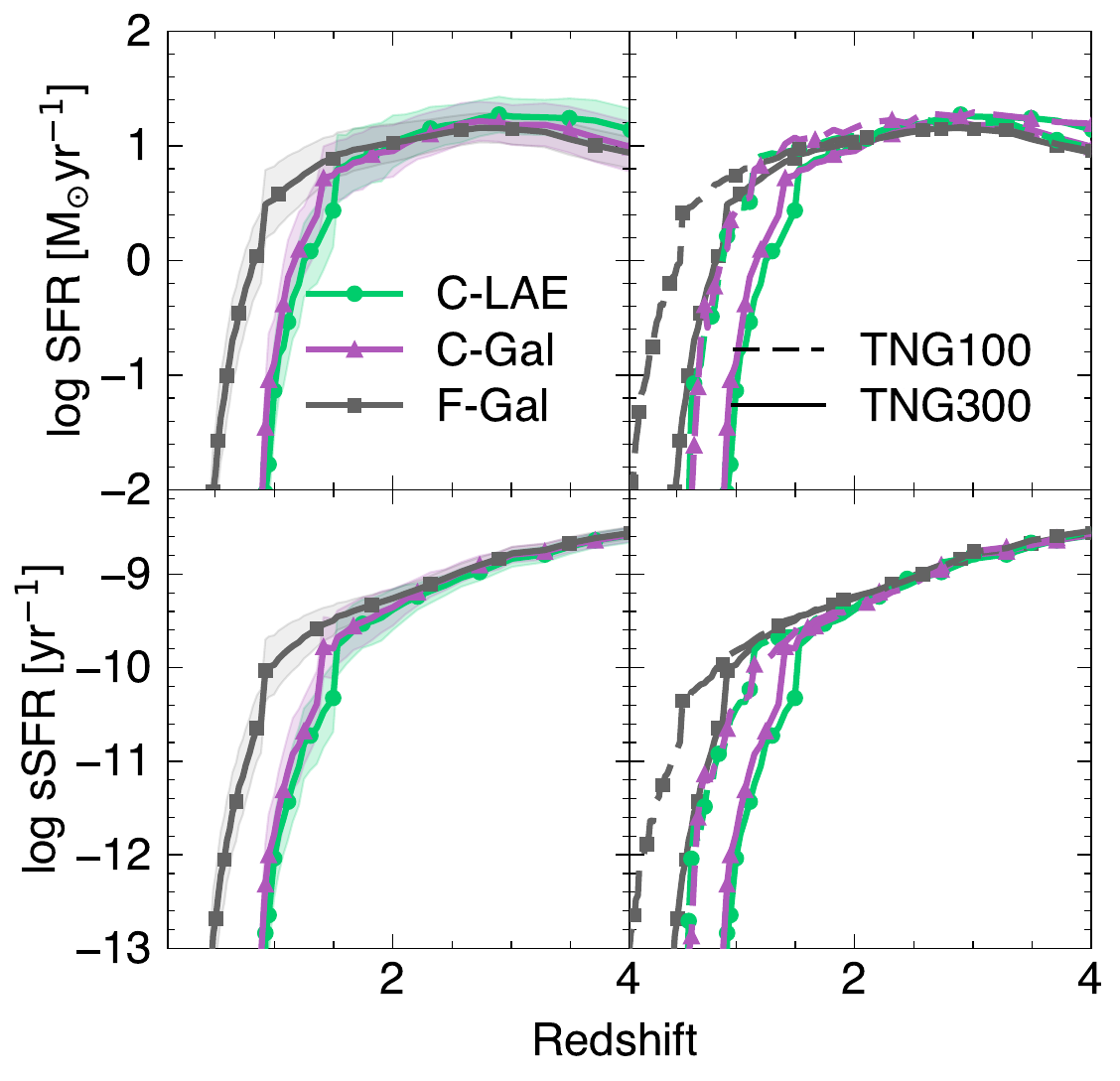}
    \includegraphics[width=.44\textwidth]{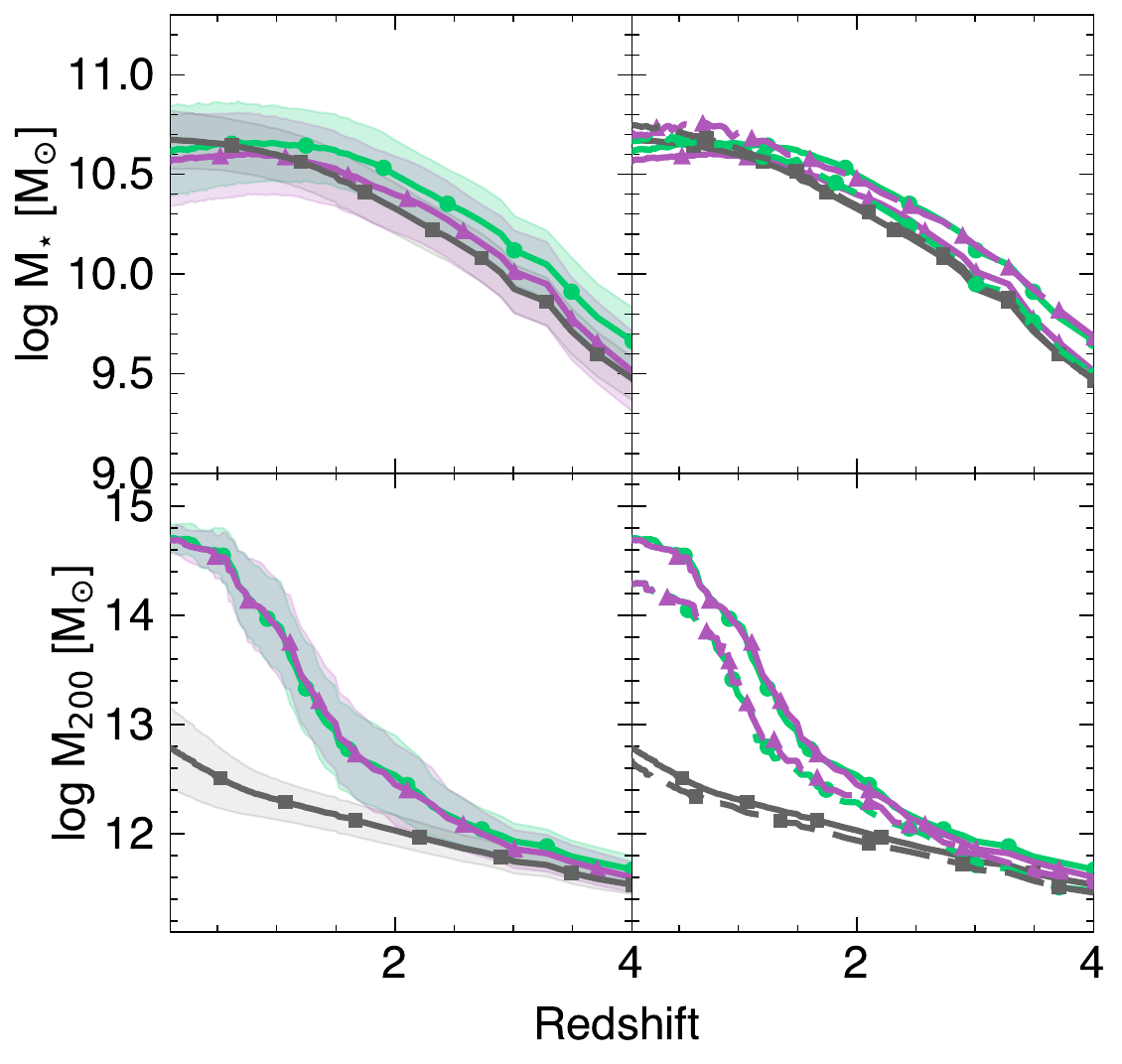}
  \caption{Cosmic evolution of SFR, M$_{\star}$, sSFR, and M$_{200}$ for the C-LAE (green lines with circles), F-Gal (gray lines with squares), and C-Gal (purple lines with triangles) samples between $z=0-4$ for TNG300 (solid lines) and TNG100 (dashed lines) simulations. Each line represents the mean of the galaxy dataset. The left panels of each block (first and third columns) show the results for only the TNG300 simulation, including errors calculated as the standard deviation. For the SFR and sSFR, the galaxies that have star formation equal to zero are included in the median calculation, but not the standard deviation (see the text for further details). The right panels of each block
(second and fourth columns) present a comparison between the results obtained for TNG100 and TNG300.}
  \label{fig:TNG300_100_comp_z4}
\end{figure*}

We now focus on the formation histories of C-Gals, i.e., galaxies that will end up as members of a massive cluster at $z=0$, in direct comparison with those of F-Gals. As a subset of C-Gals, we also examine the time evolution of C-LAEs. To do this, we use the main branch of the {\sc sublink} merger trees. For completeness, we only consider galaxies with $\Mstar \geq 10^9~\Msun$ identified at $z=4$, i.e., at the earliest snapshot we keep track of. We performed the same analysis on the galaxies identified at $z=2$ and $z=3$ and found no difference in the results that we believe would bias our results to more massive galaxies from $z=4$.

In Figure~\ref{fig:TNG300_100_comp_z4}, we show the evolution of SFR, $\Mstar$, specific SFR (${\rm sSFR}\equiv {\rm SFR}/\Mstar$), and $M_{200}$ with redshift for C-LAEs (green), C-Gals (purple), and F-Gals (gray), averaged over all galaxies in each TNG300 and TNG100 subsample. 
The first and third columns show the results for TNG300 only, with the errors calculated as the standard deviation, while the second and fourth columns show the comparison of TNG300 and TNG100.
The shaded regions are the standard deviation of star-forming galaxies at each snapshot. 
This is because the moment a galaxy becomes quiescent, the SFR drops to zero, causing a broad uncertainty margin from the median value to zero. We only consider the star-forming galaxies for error analysis to properly catch the spread of galaxies as the overall cluster SFR (and sSFR) declines.

The redshift evolution of C-Gal and C-LAE is generally similar; this is expected given that LAEs are modeled to be a subset of C-Gal albeit slightly skewed toward low-mass systems (see Section~\ref{sec:model}). Cluster galaxies experience rapid quenching commencing at $z\approx 1.6$ when SFR and sSFR values fall off steeply, while their field counterparts (F-Gal) continue to form stars until much later, at $z\approx 0.8$. While both C-Gal and F-Gal will grow to have a stellar extent of $\log \Mstar/M_\odot \sim 10.5$ by $z=0$, their large-scale environment, characterized by the host halo mass $M_{200}$, will be dramatically different. While the great majority of F-Gal will be in a small group-like environment with $\log M_{200}/M_\odot \approx 13$, the halos hosting cluster galaxies will rapidly undergo multiple mergers, growing into $\log M_{200}/M_\odot\approx 14-15$ in total mass. The slight decrease of stellar content in cluster galaxies between $z=0$ and the time of quenching is due to the mass loss expected from stellar evolution.

The second and fourth columns of Figure~\ref{fig:TNG300_100_comp_z4} compare the results obtained from TNG300 and TNG100. While the overall agreement is good in all four probed quantities, it can be seen that in TNG100, quenching occurs at a later time for both cluster and field galaxies relative to TNG300 predictions, moving from $z\approx 1.6$ to $\approx 1$ for cluster galaxies, and from $z \approx 0.8$ to $\approx 0.4$ in the average field. This difference is mainly driven by halo masses. As discussed in Section~\ref{sec:Protocluster_Selection} and listed in Table~\ref{tab:TNG_protocluster_properties}, the descendant masses ($M_{200}$) of TNG100 structures are lower compared to those in TNG300 due to its smaller volume, resulting in late quenching for their galaxy constituents. Similarly, even for the field galaxies, the host halo masses in TNG100 are lower than those in TNG300. Different resolutions of these simulations may also contribute to these variations, as the lower resolution in TNG100 may affect the treatment of small-scale processes such as gas cooling, feedback, and star formation.

\begin{figure}[h!]
    \centering
    \includegraphics[width=.4\textwidth]{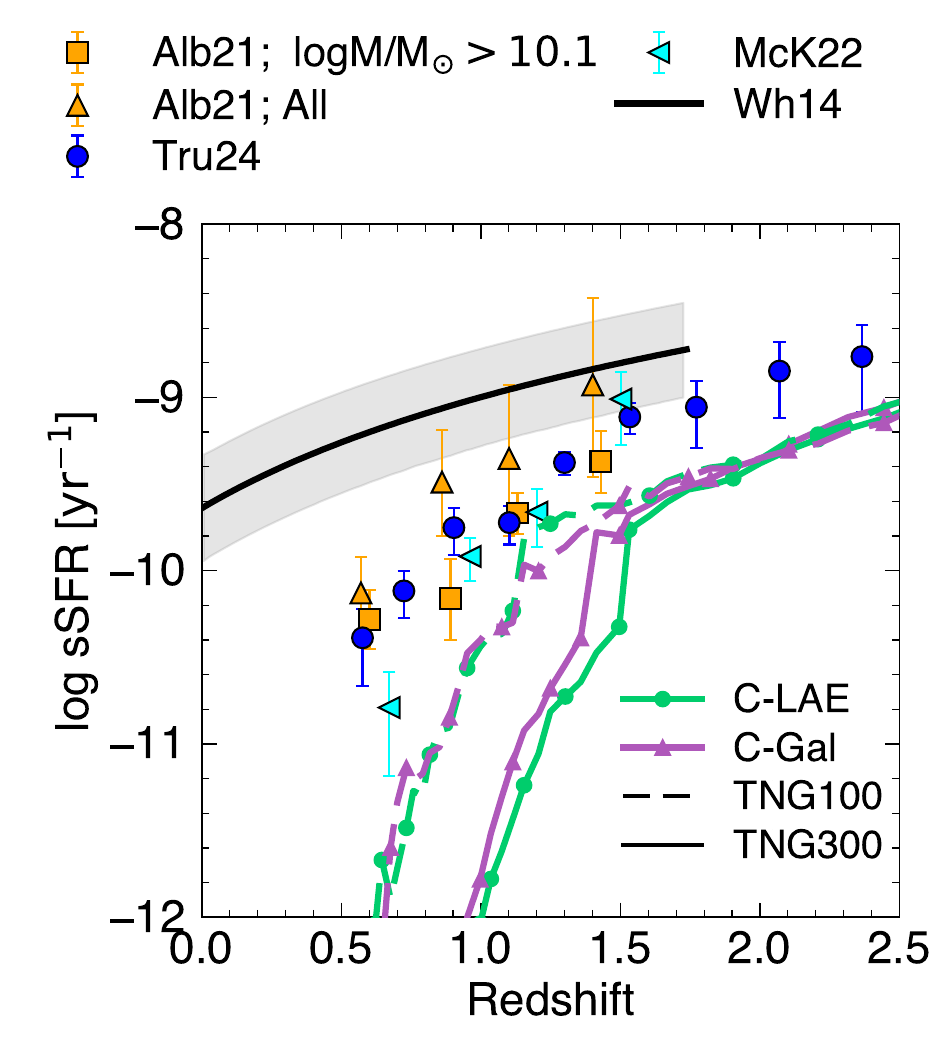}
  \caption{Comparison of the sSFR as a function of redshift of the C-LAE (green lines with circles) and C-Gal (purple lines with triangles) samples for TNG300 (solid lines) and TNG100 (dashed lines).
  We compare our results with observed cluster galaxies from \citet[][cyan left triangles]{McKinney_2022} and from \citet[][orange triangles]{Alberts2021}  using the total light from clusters. Orange squares indicate the measurements for cluster members with $\log(\Mstar / \Msun)>10.1$. We also include data from \citet[][blue circles]{trudeau24} of cluster galaxies. Gray shaded regions denote the star-forming main sequence for field galaxies at a fixed mass of $\log( \Mstar/ \Msun) = 10$, from  \citet{Whitaker_2014}.}
  \label{fig:merger_tree_obs_comp}
\end{figure}

It is challenging to confront these TNG predictions against observational data. Unlike TNG predictions that trace the same galaxies across cosmic time, 
the progenitor-descendant relationship between real clusters identified at different cosmic epochs is poorly understood. For example, based on the clustering measurements of 277 galaxy clusters identified in the IRAC Shallow Cluster Survey \citep[ISCS:][]{icsc}, \citet{brodwin07} concluded that the host halo masses at the time of observation remain similar ($\approx 10^{13.8}M_\odot$) at $z\approx 0.5$ and $z\approx 1$. The result implies that there is no direct evolutionary connection between these cluster subsamples. In the very large cluster samples assembled by \citet{trudeau24} based on the Massive and Distant Clusters of WISE Survey 2 (MaDCoWS2), which consist of over 10,000 systems at $z=0.5-2.0$, clustering measurements do not yet exist. 

\begin{figure*}[ht!]
    \centering
    \includegraphics[width=0.6\textwidth]{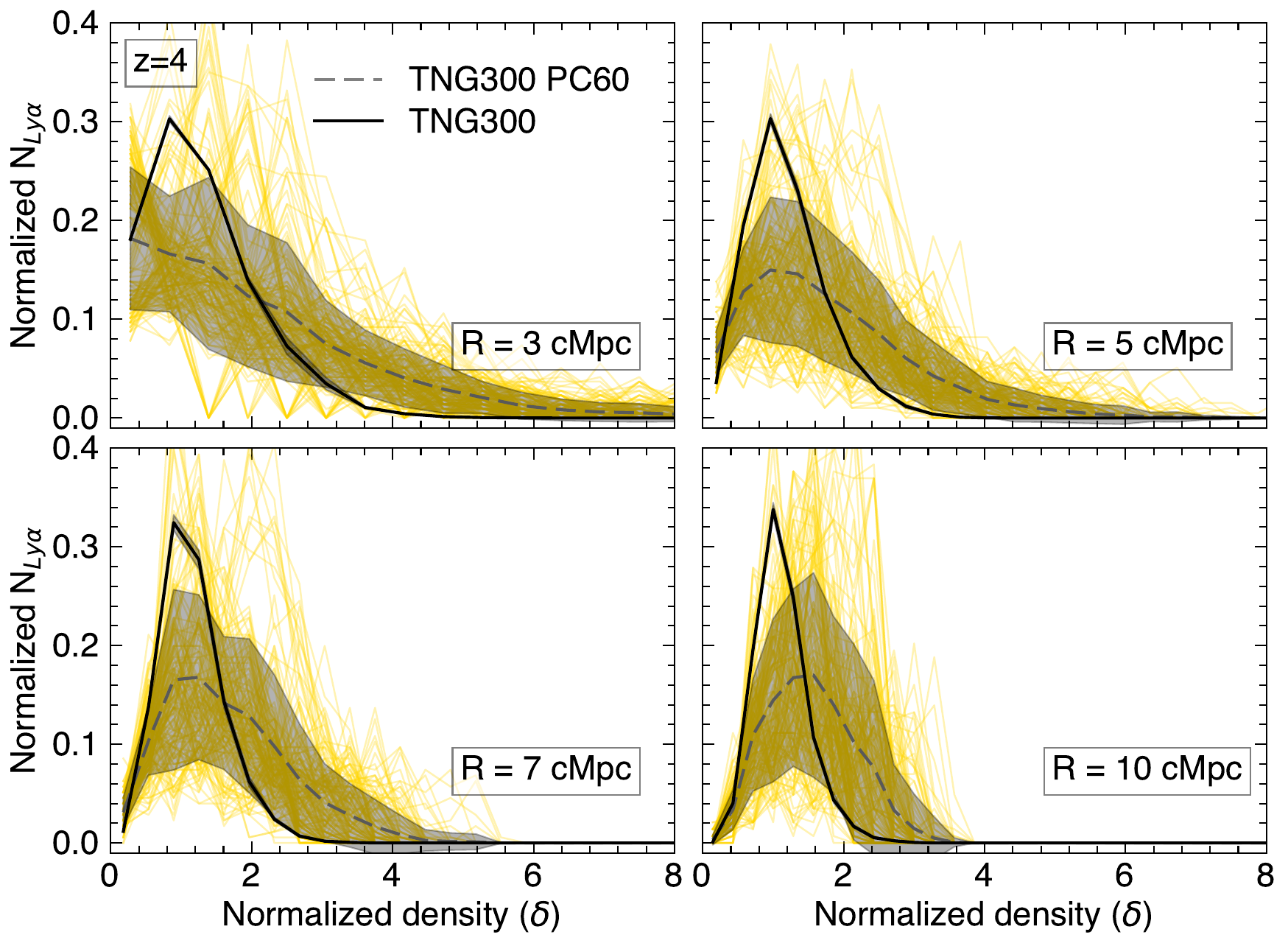}
    \caption{Distribution of the large-scale environment of the LAEs from TNG300 at $z=4$. Each panel shows the results for a different radius $r=3, 5, 7$, and 10~cMpc (from top left to bottom right). We include the mean value for the TNG Full box (solid black line), the results obtained for each PC box (solid yellow faint lines), and their average distribution (gray dashed lines).
    }
    \label{fig:env_dens_at_z_4}
\end{figure*}

Keeping in mind that direct comparisons are not straightforward, we show in Figure~\ref{fig:merger_tree_obs_comp} the sSFR as a function of redshift as measured by \citet{Alberts2021,McKinney_2022,trudeau24} together with the TNG predictions. The notations denoting different galaxy subclasses and TNG100/300 simulation are identical to those shown in Figure~\ref{fig:TNG300_100_comp_z4}.  While similar measurements are available from other cluster samples \citep[e.g.,][]{rykoff16,oguri18}, the studies featured in the figure employ an image stacking method similar to the intensity mapping technique designed to capture all cluster light from star formation activity and stellar content \citep[see][for further details]{Alberts2021} and thus are ideal for comparison. Additionally, we show the \citet{Alberts2021} measurement made only for massive galaxies ($\log \Mstar / M_{\odot}>10.1$) individually detected in the Spitzer/IRAC data as yellow triangles. For reference, we also show the star-forming main sequence at $\log(\Mstar / \Msun)=10.1$ from \citet[][the thick solid black line and the gray shaded region]{Whitaker_2014}.

As can be seen in Figures~\ref{fig:obscomp_hist_z234} and \ref{fig:merger_tree_obs_comp}, the overall normalization of the SFMS based on observations from \citet{Speagle_2014} and \citet{Whitaker_2014}  tends to be higher than the predictions of TNG\null. We refer interested readers to the discussions of possible causes for this disagreement given in \citet[][see their Section~4.3]{Donnari_2019}. 
By comparing the \citet{Whitaker_2014} SFMS with their measurements of sSFR averaged over all cluster galaxies, \citet{Alberts2021} noted that the rapid decline in star formation activity in clusters relative to the field commences at $z\approx 1.5$. Massive galaxies (yellow squares in Figure~\ref{fig:merger_tree_obs_comp}), while having lower sSFR values, quench alongside all galaxies dominated by their lower-mass cousins. The \citet{McKinney_2022} results (cyan squares), an update to the \citet{Alberts2021} with improved stellar mass constraints, as well as those reported by \citet[][blue circles]{trudeau24} also paint a consistent picture, lending credence to rapid quenching in cluster galaxies relative to the average field galaxies. 

The difference between TNG300 and TNG100 in their average formation histories may be in part explained by the fact that TNG300 contains more massive clusters for which quenching occurs at an earlier time than those in TNG100. Thus, the slower rate of decline in sSFR observed for the ISCS and MaDCoWS cluster members may be interpreted as a superposition of galaxy clusters of decreasing descendant masses undergoing quenching at increasingly later times. Thus, the observed decline in star formation activity in clusters is qualitatively in agreement with the TNG predictions.

\section{Discussion: The prospect for the ODIN survey}\label{sec:discussion}

As the widest-field deep narrowband survey to date, the ODIN survey is expected to identify over 100,000 LAEs within a total comoving volume of $\approx 2\times 10^8$~cMpc$^3$ at Cosmic Noon \citep{Lee2024}. One of the scientific goals of ODIN is to identify the sites of hundreds of massive protoclusters via their significant LAE (surface) overdensities and examine how galaxy formation is linked to the surrounding large-scale structure \citep[e.g., see][]{Ramakrishnan_2023}. 
In the first protocluster study based on ODIN, \citet{ramakrishnan24} showed that the observed protoclusters will likely evolve to Virgo analogs with masses of $\log (M_{z=0}/\Msun)\sim 14.0 - 15.0$. Cosmic structures in TNG and their galaxy constituents considered in the present study are well-matched and thus can inform future galaxy evolution studies to be conducted by ODIN.

The narrow-band-based galaxy selection, adopted by ODIN and several other existing surveys, is expected to yield robust samples of protoclusters with a negligible and well-quantified contamination fraction \citep{ramakrishnan24}. 
Still, even with extensive spectroscopic follow-up efforts, the precise membership of individual galaxies will be out of reach for most protoclusters. In this section, we implement observational limitations into the TNG simulations to develop a realistic sense of what galaxies in a protocluster field may look like in the real data.

\begin{figure}[h!]
    \centering
    \includegraphics[width=\columnwidth]{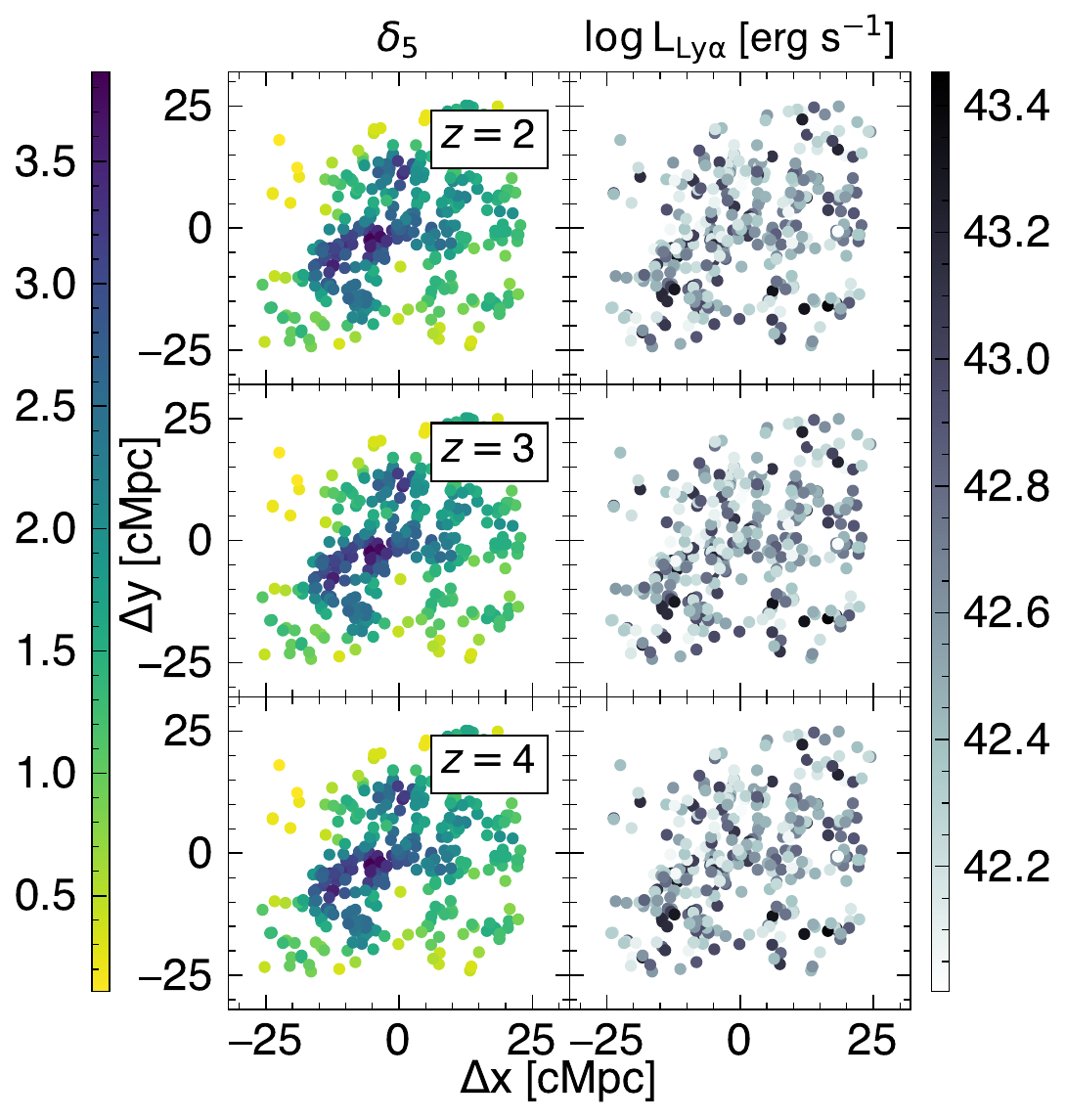}
    \caption{Spatial distribution of LAEs in the XY- plane for the region centered in the most massive cluster selected at $z=0$ from TNG300, and traced through $z= 2$ (top), 3 (middle), and 4 (bottom). The left panels display the LAE positions colored by environmental density $\delta_{5}$ while the right panels show the same positions colored by the Lyman $\alpha$ luminosity.}
    \label{fig:X_Y_MassiveProtocluster_z234}
\end{figure}

The ODIN filters are 70--100~\AA\ wide, corresponding to the line-of-sight thickness of 50--75~cMpc. To match this, we use the PC60 boxes (60~cMpc wide on a side) centered around the 30 most massive clusters. We collapse them along the X, Y, or Z direction, thus producing `2D images' of 90 ($=30\times 3$) protocluster sightlines. These boxes correspond to the transverse size of $\approx$27\arcmin--34\arcmin, leaving a large non-protocluster area flanking each protocluster in the center. To compute the normalized local density of a given LAE, $\delta$, we first count the number of neighbor LAEs within a circular aperture of radii $r_i$ (3, 5, 7, and 10~cMpc), $N_i$. We use the full TNG volumes, similarly sliced to a 60~cMpc thickness, and calculate the corresponding field average as $N_{{\rm avg},i} = \bar{\Sigma}_{\rm LAE} \cdot \pi r_i^2$. The normalized local density is then simply $\delta_i \equiv N_i/N_{{\rm avg},i}$. The adopted method is similar to the smoothed density maps used by recent protocluster studies \citep[e.g.,][]{badescu17,shi19,forrest23,Ramakrishnan_2023} where $r_i$ signifies the smoothing scale. For reference, many of these existing studies typically employed Gaussian smoothing with a half-width-at-half-maximum of $\approx$5~cMpc. Our method of computing 2D surface density should realistically simulate ODIN-derived densities where unassociated foreground and background interlopers dilute the true density fluctuations across the field.

Figure~\ref{fig:env_dens_at_z_4} shows the distribution of $\delta$ values averaged over all $z=4$ LAEs in PC60 volumes as thick dashed lines. Individual distributions of 90 PC60 boxes are shown in yellow. The thick black lines show the expectation in a (60~cMpc)$^3$ volume, estimated from the full TNG simulation. All distributions are normalized such that the area under it is equal. As expected, the $\delta$ distribution in average fields always peaks at $\delta\approx 1$ regardless of the $r_i$ values. Larger smoothing scales, $r_i$, tend to make the overall $\delta$ distribution narrower and move closer to $\delta=1$. The distribution is much broader in protocluster volumes and shows a long tail on the high $\delta$ side. This is true even though the transverse area we chose is much larger than the typical size of a protocluster core, which is no more than a few arcminutes across\footnote{Assuming the Planck cosmology, 1~arcminute corresponds to 1.54, 1.89, and 2.14~cMpc at $z=2$, 3, and 4, respectively.}.  For the $\delta_5$ (density at $r=5$~cMpc) distribution, for example, a 28\arcmin$\times$28\arcmin\ field centered on a $z=4$ protocluster, approximately one-third of all LAEs would reside in a region whose density is at least twice the field average or greater. Additional 10-15\% of the LAEs, populated over one-quarter of the total area, live in densities well below the average. Our finding suggests that wide-area coverage extending a few degrees on a side is essential to estimate true average galaxy density thereby robustly detecting LSS features as surface overdensities of LAEs.

The left panels of Figure~\ref{fig:X_Y_MassiveProtocluster_z234} present the 2D spatial distribution of the LAEs at $z=2, 3$ and 4 for the PC60 box corresponding to the most massive cluster identified at $z=0$ for TNG300 color-coded by $\delta_{5}$. As expected, LAEs near the core have the highest $\delta_{5}$ values.
Similar results are obtained for $\delta_{3}$, $\delta_{7}$, and $\delta_{10}$ (density at $r=3$, 7, and 10~cMpc, respectively). The right panels show the same 2D map but this time color-coded by Ly$\alpha$ luminosities. No clear correlation between $\delta_{5}$ and $L_{{\rm Ly}\alpha}$ is found. In other words,  the brightest LAEs in TNG simulations are not preferentially placed in environments with the highest density. While this is in line with the fact that the protocluster LALF is rather similar to that in an average field other than the higher overall normalization, as discussed in Section~\ref{subsec:pclf} and shown in the right panels of Figure~\ref{fig:LF_LAE_Protoclusters}, it is at odds with the result from \citet{dey16}, who reported a decrease in line luminosity with increasing distance to the protocluster center around several LAE-selected protoclusters at $z=3.78$ (see their Figures 10 and 11). With large, robust samples of protoclusters, ODIN will unambiguously determine how line luminosities and equivalent widths may change in protocluster environment \citep[e.g.,][]{dey16,lemaux18,lemaux22}.

\begin{figure}
    \centering  
    \includegraphics[width=1.05\columnwidth]{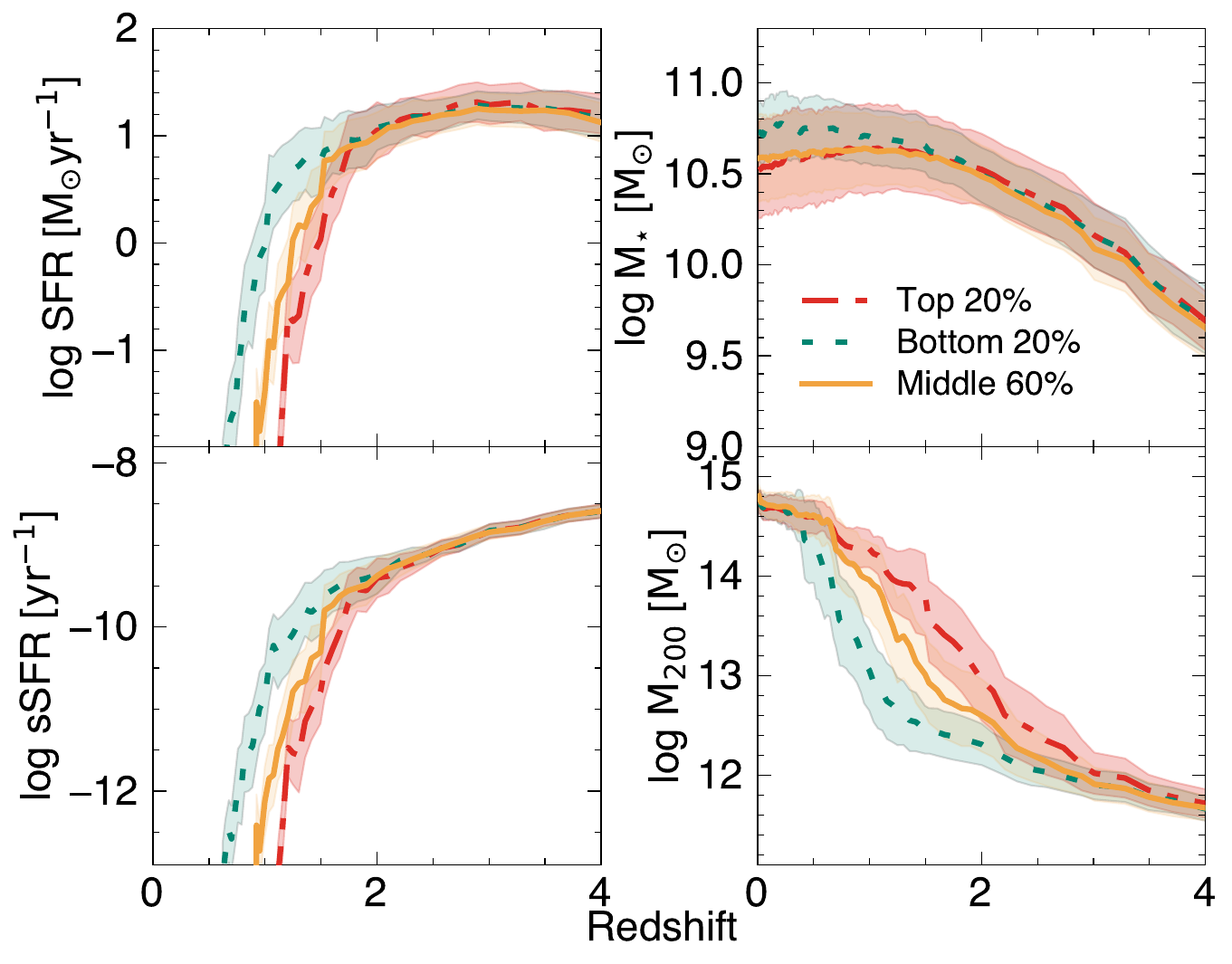}
  \caption{Cosmic evolution of SFR, $M_{\star}$, sSFR, and $M_{200}$ of TNG300 galaxies are compared for C-LAEs split by environmental density $\delta_5$. The C-LAEs in the top 20\% densest regions are shown in red (dot-dashed line), the middle 60\% are shown in orange (solid line), and the bottom 20\% are shown in blue (dotted line). 
  The averages and standard deviation are computed the same as for Figure~\ref{fig:TNG300_100_comp_z4}. 
  }
  \label{fig:TNG300_100_LAEcomp_z4}
\end{figure}

Finally, we examine the evolution of LAEs as a function of their 2D environment. To this end, we split the C-LAEs into three bins: the top 20\%, bottom 20\%, and the middle 60\% in their measured $\delta_5$ values, and track these groups across cosmic time from $z=4$ to $z=0$ using the same methodology as described in Section~\ref{sec:evolutionLAEs}. The results for TNG300 are shown in Figure~\ref{fig:TNG300_100_LAEcomp_z4}. The galaxies in the top 20\% group experience the earlier decline in SFR and sSFR and belong to more massive halos (${\rm M_{200}}$) than the bottom 20\% group. Since the top 20\% region should lie closest to protocluster cores, the implication is that galaxy evolution in clusters is an inside-out process in which quenching occurs in the densest knots and spread to lower-density regions and to galaxies in lower-mass halos. This is consistent with other studies using the Horizon Run 5 simulation \citep{Leesk_2024}.

Figure~\ref{fig:TNG300_100_LAEcomp_z4} projects an optimistic forecast for future observational studies.  The clear $\delta$-dependent evolutionary sequence demonstrates that the 2D environment defined by ODIN should do an adequate job pinpointing the key features of the underlying LSS including the densest cores and their outskirts, despite the absence of the line-of-sight (redshift) information. In other words, it may be sufficient to limit any searches for galaxies that are quiescent or currently being quenched to regions of the highest LAE surface densities.

\section{Summary and conclusions}\label{sec:conclusion}

In this study, we examine the galaxy properties in high-redshift protocluster and their evolution using IllustrisTNG. We define protoclusters as large cosmic structures that will evolve into the most massive clusters at $z=0$. To build clear and realistic expectations for ODIN, we select the 30 (10) most massive halos of TNG300 (TNG100) ranked by their stellar mass at $z=0$ and create subvolumes centered around the barycenters. We map the galaxy properties within these sub-volumes to $z = 2, 3,$ and 4. The robustness of our findings is tested by comparing the results from TNG300 and TNG100 runs, which cover different cosmic volumes and resolutions but share the same subgrid physics. 
Our findings are summarized as follows.

\begin{enumerate}
    
    \item We find a clear difference in the overall shape and amplitude of the UV luminosity function of galaxies in protocluster- and field environments at $z=2-4$. In a (15~cMpc)$^3$ volume centered on protoclusters, the UVLF is characterized by the faint-end slope that is shallower by $\Delta \alpha \approx 0.3$ relative to the field UVLF as well as a significant excess on the bright end (Figure~\ref{fig:LF_UV_Protoclusters}, Table~\ref{tab:UVLF}). These findings suggest that the formation history of protocluster galaxies may be markedly different compared to those in the field. If we confine our measurement to regions exhibiting a high galaxy overdensity (instead of identifying them at $z=0$ and tracing their properties back to high redshift), the protocluster-to-field excess at the bright end is even more pronounced ($\approx 30$ for galaxies with $-20 < M_{\rm UV} < -23$, see Figure~\ref{fig:UVLF_for_fornax_PC}).\\

    \item The difference in the overall shape of the Ly$\alpha$ luminosity function in protocluster- and field environments is subtler. While the LALF within a protocluster volume exhibits an overall higher normalization, there is no clear excess of galaxies on the bright end nor is the difference in the faint-end slope severe (Figure~\ref{fig:LF_LAE_Protoclusters}). This, combined with the fact that the bright end of measured LALFs is almost always contaminated by AGN, suggests that LALFs may not be as sensitive a tool as UVLFs in investigating galaxy formation as a function of environment. \\

    \item Protocluster member galaxies obey the same SFMS on the SFR-$\Mstar$ plane as the average field galaxies (Figure~\ref{fig:obscomp_hist_z234}). Given the different shapes of the UVLFs and LALFs, protocluster galaxies may slide along the star-forming main sequence toward the high SFR and high $\Mstar$ end while steadily growing at higher rates than non-protocluster galaxies. The implication is that, even in dense protocluster environments, only a small fraction of galaxies may be dominated by bursty star formation, which would place them above the SFMS. \\

    \item Galaxies that underwent major mergers in the last 200~Myr with a mass ratio of 1:$n$ ($n\leq 4$)  experience 
    an increase of their star formation with SFR enhanced at a $\approx 60\%$ level at a fixed stellar mass. While the increase in SFR  uniformly applies in any environment, the higher merger rates and the higher number of massive halos in protocluster environment lead to a flatter distribution in both SFR and $\Mstar$, characterized by a significant excess on the high-SFR/$\Mstar$ end and a deficit on the low end, as illustrated in Figure~\ref{fig:merger_comp}. The difference in the shape of the protocluster- and field UVLFs (Sections~\ref{subsec:LF_field} and \ref{subsec:pclf} and Figure~\ref{fig:LF_UV_Protoclusters}) is primarily attributed to galaxy mergers. \\
        
    \item While galaxies identified at $z=4$ evolve to similarly massive galaxies by $z=0$, they undergo distinctly different formation histories. Protocluster member galaxies, including LAEs, begin to quench at $z\approx 1.6$ well before the field counterparts do (Figure~\ref{fig:TNG300_100_comp_z4}, Section~\ref{sec:evolutionLAEs}). While the precise onset of quenching in clusters remains uncertain and resolution-dependent, the time sequence of quenching as measured by sSFR versus redshift is in qualitative agreement with existing measurements of cluster galaxies (Figure~\ref{fig:merger_tree_obs_comp}). Galaxy evolution in clusters appears to occur inside out, with quenching occurring to the galaxies in the highest-density protocluster cores first then extending to lower-density regions and lower-mass halos later (Figure~\ref{fig:TNG300_100_LAEcomp_z4}). \\
    
    \item Using subvolumes matched to the ODIN survey geometry, we simulate what real data in protocluster fields would appear and compute local (surface) density, $\delta$, at scales of 3, 5, 7, and 10~cMpc from each LAE. 
    Our results show that, in protocluster volumes, the distribution $\delta$ deviates significantly from the global average, showing a long tail on the high $\delta$ side in all scales. 
    To accurately estimate local density and detect LSS features accordingly, wide-field imaging of a few degrees on a side would be required (Figures~\ref{fig:env_dens_at_z_4} and \ref{fig:X_Y_MassiveProtocluster_z234}).
    Additionally, despite the relatively large line-of-sight thicknesses employed by narrow-band surveys, 2D-based environmental density estimates adequately pinpoint the key regions of protoclusters including their approximate centers (Section~\ref{sec:discussion}). 
\end{enumerate}

Finally, our study highlights significant differences in the galaxy properties and their evolution in protoclusters compared to the field. Observational efforts, including ODIN and future LSST and Roman, will bring into sharp focus how diverse cosmic environments drive the formation and evolution of galaxies.

\begin{acknowledgements}
We thank the referee for her/his useful comments. 
MCA and AK acknowledge the financial support from 
ANID/Fondo 2022 ALMA/31220021. MCA acknowledges support from ANID BASAL project FB210003. KSL and VR acknowledge financial support from the National Science Foundation under Grant Nos. AST-2206705 and AST-2408359, and from the Ross-Lynn Purdue Research Foundation. EG acknowledges support from the National Science Foundation Grant AST-2206222.
The Institute for Gravitation and the Cosmos is supported by the Eberly College of Science and the Office of the Senior Vice President for Research at the Pennsylvania State University.
JL is supported by the National Research Foundation of Korea (NRF-2021R1C1C2011626).
We acknowledge the computational resources provided by the Rosen Center for Advance Computing at Purdue University and the computer server RAGNAR at Universidad Andres Bello. We also made partial use of the FASRC Cannon cluster supported by the FAS Division of Science Research Computing Group at Harvad University. 
\end{acknowledgements}

\bibliographystyle{aa}
\bibliography{manuscript}

\end{document}